\documentclass{article}
\usepackage[utf8]{inputenc}
\usepackage[T1]{fontenc}
\usepackage[title]{appendix}%

\newcommand{\keepcomment}{0} % 1 - Keep comments, 0 - Hide comments
\usepackage[normalem]{ulem}
\ifnum\keepcomment=1
	\usepackage[colorinlistoftodos,textsize=scriptsize]{todonotes} % todo
    \newcommand{\stkout}[1]{\ifmmode\text{\sout{\ensuremath{#1}}}\else\sout{#1}\fi}
    \newcommand\rev[3]{\textcolor{red}{\begin{scriptsize}{#1}\end{scriptsize}\stkout{#2}}\textcolor{blue}{#3}}
\else
	\newcommand\rev[3]{\ignorespaces#3\ignorespaces\unskip}
	\usepackage[disable]{todonotes} % todo
\fi

\usepackage{amsmath}
\usepackage{amsfonts} % For mathbb

\usepackage{calrsfs} % For pazocal
\DeclareMathAlphabet{\pazocal}{OMS}{zplm}{m}{n}

\usepackage{hyperref}
\hypersetup{
    colorlinks=true,
    linkcolor=blue,
    filecolor=magenta,      
    urlcolor=cyan,
    pdftitle={Overleaf Example},
    }

\usepackage[margin=1in]{geometry}

\title{Quantifying the Improvement of Accessibility achieved via Shared Mobility on Demand}

\author{Severin Diepolder 
\\ Wavestone, Germany
\and
Andrea Araldo
\\ 
SAMOVAR, Télécom SudParis, Institut Polytechnique de Paris, 91120 Palaiseau, France
\and
Tarek Chouaki
\\IRT System X, France
\and
Santa Maiti
\\University College London, United Kingdom
\and
Sebastian Hörl
\\IRT System X, France
\and
Constantinos Antoniou
\\Chair of Transportation Systems Engineering, Technical University of Munich, Germany}

\begin{document}

\maketitle

\begin{abstract}
Shared Mobility Services (SMS), e.g., demand-responsive transport or ride-sharing, can improve mobility in low-density areas, which are often poorly served by conventional Public Transport (PT). Such improvement is generally measured via basic performance indicators, such as waiting or travel time. However, such basic indicators do not account for the most important contribution that SMS can provide to territories, i.e., increasing the potential, for users, to reach surrounding opportunities, such as jobs, schools, businesses, etc. Such potential can be measured by \emph{isochrone-based accessibility} indicators, which count the number of opportunities reachable in a limited time, and are thus easy for the public to understand.
The potential impact of SMS on accessibility has been qualitatively discussed and implications on equity have been empirically studied. However, to date, there are no quantitative methods to compute isochrone-based  indicators of the accessibility achieved via SMS.

This work fills this gap by proposing a first method to compute isochrone accessibility of PT systems composed of conventional PT and SMS, acting as a feeder for access and egress trips to/from PT hubs. This method is grounded on spatial-temporal statistical analysis, performed via Kriging. It takes as input observed trips of SMS and summarizes them in a graph. On such a graph, isochrone accessibility indicators are computed. We apply the proposed method to a MATSim simulation study concerning demand-responsive transport integrated into PT, in the suburban area of Paris-Saclay.

\end{abstract}

% suggestion Santa: aa:Integrated
%\textcolor{blue}{Shared Mobility Services (SMS), e.g., Demand-Responsive Transit (DRT) or ride-sharing, can improve mobility in low-density areas, often poorly served by conventional Public Transport (PT). To indicate the improvement, most of the works used basic performance indicators, like wait or travel time. However, accessibility represents ease of reaching opportunities is more comprehensive than basic indicators. A method that quantifies DRT'\mathbf{s} impact on accessibility is missing in the literature. The main challenge in quantification is accessibility is generally computed on graph representations of PT networks. However, SMS trajectories are dynamic and do not follow a predefined network. We propose a spatial-temporal statistical method that summarizes observed SMS trips in a graph on which accessibility can be computed. We apply our method to a MATSim simulation study concerning DRT in Paris-Saclay.}


\section{Introduction}
Location-based accessibility measures the ease of reaching surrounding opportunities, such as education, health, cultural or shopping places, via transport (\cite{miller2020accessibility}). The ease is mainly measured in terms of time needed, but it can also include other factors, such as monetary cost, discomfort. Accessibility provided by conventional Public Transport (PT) is generally poor in low-demand areas, e.g., suburbs (\cite{badeanlou2022ptanalysistool}), because a high-frequency and high coverage service in such areas would imply an unaffordable cost per passenger. Poor PT accessibility in the suburbs makes them car-dependent, which prevents territories from being sustainable (\cite[Section~2.2]{Boussauw2022}). Car dependency undermines the three pillars of sustainability: (i) Environment: 61\% of carbon dioxide~(CO\textsuperscript{2}) emitted by European road transport comes from cars (\cite{EUFacts2022}), (ii) Society: people without cars can have up to 5 times less access to jobs (\cite{Bertolini2021}) and (iii)~Economy: the time spent in congestion amounts to 100 billion~€ in Europe (\cite{EUTransport2020}). The lack of satisfying accessibility from PT, and the consequent car dependency, might explain why the suburban population pollutes on average more than those living in the centre. Such fact has been observed by \cite{Pottier2020} (Figure~4) in France.

Shared Mobility Services (SMS), e.g., Demand-Responsive Transit (DRT), ride-sharing, carpooling, car-sharing, are potentially more efficient than conventional PT in the suburbs (\cite{Calabro2021}). However, their current deployment is often led by private companies targeting profit maximization. This may turn SMS into a competitor of sustainable modes, such as Public Transport (PT) (\cite{Cats2022}), and an additional source of congestion and pollution (\cite{Marshall2019, Erhardt2019a}). Such negative effects could be mitigated if SMS is integrated with conventional PT, i.e., if SMS and PT are planned and operated together, aimed at some global objective, as advocated by \cite{Fielbaum2024}, \cite{Calabro2021}. Such \emph{multimodal PT}, integrating both conventional PT and SMS, has been largely studied in work concerning their planning, generally aimed at minimizing some generalized cost (\cite{Fielbaum2024}, \cite{Calabro2021}).  However, such studies neglect accessibility, which is however \rev{}{}{today often considered} the most important objective we can aim to satisfy via a transport system.\footnote{\rev{}{}{Here, we adopt the viewpoint of \cite{miller2018accessibility}: \emph{``the primary role of a transport system is to provide people and businesses with access to other people and businesses so that they can physically engage in spatially and temporally distributed activities of all kinds, and so that they can physically exchange information, goods and services''}.}}
We argue that the benefits of integrating SMS and PT should be studied under the lens of accessibility. To this aim, there is a need to assess the accessibility resulting from such an integration, and to use such an assessment to guide deployment decisions.
However, no quantification methods exist to measure the accessibility achieved by integrating SMS and~PT. 

 In this paper, we devise a method to objectively quantify the impact of SMS on accessibility, when SMS is added as a feeder of conventional PT, e.g., it provides first and last mile connection to PT hubs.
The type of SMS on which we focus in this paper is a Demand-Responsive Transit~(DRT), although we also discuss how to adapt it to other types of SMS. 
Our method takes empirically observed trips as input. Developing such a method is challenging, because most accessibility definitions are based on computing shortest paths on a graph. While a graph (in particular a time-expended graph) can model well conventional PT (\cite{fortin2016innovative}) and road networks, SMS do not have a natural graph representation. Indeed, the routes of SMS are stochastic and change over time. As a consequence, it would be misleading to compute accessibility on a graph trivially composed of connections served by observed trips (for instance trips of the day  before). Such a computation would just give us an idea of what opportunities people reached in the past and how easy it was for them. This violates the spirit of accessibility indicators, which aim instead to tell the \emph{potential} of reaching opportunities, even those that have not been accessed in the recent past.
Computing accessibility from SMS thus requires to extract, from past observed trips, such a potential. This calls for statistical methods that can ``summarize'' observations (i.e., performance of observed trips) over multiple days and origin-destination pairs, so as to infer a general potential of reaching opportunities.

The novel contribution of this paper consists in developing a spatial-temporal statistical method to transform a dataset of previously observed SMS trips in a graph representation, on top of which well-established accessibility computation can be performed. 
The observations that can be taken as input might come from a service deployed in the real world or from a simulated service. This paper's observations come from a MATSim simulation of a DRT, acting as feeder for main metro lines in the Paris-Saclay area (\cite{Chouaki2023}). Our method is thus adapted to the case of a many-to-one and one-to-many transport service acting as a feeder for conventional PT. However, our method could later be adapted to ride-hailing, carpooling, car-sharing, bike-sharing or other shared mobility systems.

It is well known that SMS implies a high cost for the operator (\cite{Calabro2021}), and thus for the communities who might consider developing it. Therefore, such communities need to know what are the benefits of SMS. However, SMS are often evaluated in terms of level of service (ratio of requests that can be served, average trip time, etc.). By doing so, the benefits of SMS are understated, and are limited to the performance of SMS itself. The high operational cost, together with a lack of understanding and quantification of the real benefits of SMS, are barriers to their large adoption.
We argue that the real potential of SMS resides in its ability to improve accessibility in places where conventional PT cannot do it. We argue that transport planners and operators who wish to get support from the public, should be able to objectively quantify claims such as ``a certain SMS deployment gives to suburban people access to additional $x$ thousands jobs in half an hour compared to conventional PT only''. Such claims are much more meaningful than just quantifying average trip times or waiting times. We argue that such claims are crucial to foster adoption of SMS. This work provides a first method to substantiate such types of claims.

The rest of the paper is organized as follows. Section~\ref{sec:related_work} frames our contribution within the state of the art. Section~\ref{sec:methodology} is the core of the paper and describes the proposed method. Section~\ref{sec:implementation} provides some implementation details, which ensure repeatability. In section~\ref{sec:results}, we showcase our method in a case study, obtained by simulating in MATSim a perspective DRT service in the Paris-Saclay area. Section~\ref{sec:discussion_and_perspectives} discusses the limitation of the proposed method and the direction for future work. Section~\ref{sec:conclusion} concludes the paper.

Our code is available as open source (\cite{githubrepo}).

%In this sense, by providing a first method to quantify the accessibility of SMS on empirical observations, this work can contribute to foster adoption of SMS and to guide their deployment under the lens of accessibility.
%By providing a first method to compute accessibility of SMS on empirical observations, this work can contribute to better understand the potential of SMS and guide their future deployment.

%%%%%%%%%%%%%%%%%%%%%%%%%%%%%%%%%%%%%%%%%%%%%
%%%%%%%%%%%%%%%%% RELATED WORK %%%%%%%%%%%%%%
%%%%%%%%%%%%%%%%%%%%%%%%%%%%%%%%%%%%%%%%%%%%%
\section{Related Work}
\label{sec:related_work}

There exists multiple decades of excellent work regarding Shared Mobility Services (SMS) and regarding accessibility. A recent review on modelling and optimization of SMS, with a focus on Demand-Responsive Transit (DRT), can be found in Sec.~2 of \cite{Calabro2021} and in \cite{Zhu02052023}. 

As for accessibility, a short and excellent review is provided by \cite{miller2020accessibility}. Tools to evaluate accessibility are presented by \cite{Bertolini2019,Byrd2023,biazzo2019accessibility} and compared by~\cite{Miller2022}. Example of tools, already well-known by transport planners, are Conveyal Analysis (\cite{Conway2017,Conway2018}) and Remix (\cite{Laquidara2024}).

Surveying the contributions to the SMS domain and the accessibility domain separately would be out of scope and has already abundantly done by the aforementioned references. We will not repeat it here.
Within the focus of this paper, it suffices to observe that work on SMS does not generally consider accessibility and work on accessibility does not generally consider SMS, with a few nuanced exceptions, on which this section will focus.

\cite{chandra2013accessibility} studied how SMS (in particular DRT in their work) improves connection to conventional PT stops. The observation was confirmed by \cite{Calabro2021} (Figure~7), who showed that, by deploying SMS-based feeder (DRT also in this work) in the suburbs, the time to access main Public Transport (PT) corridors is reduced compared to conventional fixed-route feeder lines. However, the ``access to PT'' observed in the aforementioned two papers needs not to be confused with accessibility. Indeed, accessibility expresses the potential to access opportunities, rather than the potential to access PT stops (\cite{miller2020accessibility}). In other words, when computing accessibility, PT should not be treated as the target, but as a means. This is why, in our work, accessibility is equal to the number of opportunities that can be reached in a certain time \emph{via} SMS combined with PT.

Considerations about the impact of SMS on accessibility have been triggered by the advent of automated vehicles. \cite{VanWee2018} proposed a conceptual method, based on interviews with experts, to qualitatively evaluate the impact of automated SMS, in the form of shared taxis, on accessibility. \cite{Richter2021} proposed another conceptual method to identify the areas in which the accessibility currently provided by conventional PT is much lower than cars. Such areas were identified as the ones in which the development of automated SMS could be the most beneficial. These articles, however, were focused on qualitative considerations and do not compute the accessibility achieved via SMS, which is instead the focus of our work.

Recent research has focused on equity implications of SMS (shared taxis, bike sharing, e-scooters), studying empirical mobility data to discover how SMS usage is impacted by social, economic, demographic and geographic factors (\cite{Jiao2021,Ettema2024}). However, \cite{Jiao2021} did not refer to accessibility. \cite{Ettema2024}, instead, due to the lack of quantitative accessibility indicators for SMS, resorted to a proxy, namely the ``perceived accessibility'', obtained by collecting answers from surveys, where people were asked to declare if they had the impression they could reach more easily the gym, healthcare, supermarkets or their workplace, thanks to SMS. Our work provides a method to compute objective accessibility indicators, which go beyond users' perceptions, and allow complementary findings.

Quantification of accessibility provided by SMS (automated shared taxis, in particular) was performed by \cite{nahmias2021drtaccessibility}, \cite{zhou2021simulating} and \cite{Ziemke2023}. They all resort to random utility-based measures of accessibility (following the terminology of \cite{miller2020accessibility}). Such measures, often referred to as logsum-based, are based on utilities perceived by agents within high-detail simulations. However, random utility-based accessibility indicators have several disadvantages (\cite{GeursVanWee2013,GeursVanWee2023}). Indeed, the complexity of the development of such simulations limits the applicability of such methods. Moreover, the validity of the obtained accessibility measure is tightly dependent on the correctness of the simulation model and of its calibration, which are difficult to achieve. Furthermore, such indicators, of econometric nature, are not easily interpretable by a large audience.
For these reasons, we adopt instead an isochrone-based measure of accessibility. Such a measure is location-based: given a certain location, it counts the number of opportunities that can be reached in a given amount of time, departing from that location. This measure is easily interpretable and effectively communicated to the public and does not strictly require the development of simulation tools.

We now emphasize the difference between our work and the closest work, namely of \cite{Ziemke2023} and \cite{abouelela2024we}. \cite{Ziemke2023} computed accessibility of a taxi-like service, with one passenger at a time, going from an origin to a destination. \cite{abouelela2024we} analysed the accessibility generated by shared e-scooters, which, obviously, can be used by one user at a time. In both cases, there are no shared trips. In this work, we compute instead the accessibility of a shared feeder service, where (i)~having multiple passengers at a time is crucial for achieving the required cost-efficiency and (ii)~the accessibility gained by passengers is achieved by bringing passengers to main PT hubs from which opportunities are easily reachable. The two aspects that distinguish our work from \cite{Ziemke2023} and \cite{abouelela2024we} are thus (i)~trip sharing and (ii)~multimodality. These two aspects prevent us from computing travel times as simply inversely proportional to the speed of the considered mode (e.g., e-scooter speed, as in \cite{abouelela2024we}). The novel statistical approach we devise in this paper allows to naturally deal with these two aspects, by estimating resulting travel times from a dataset of observed trips.

Recent work focused on the accessibility improvement achieved via a feeder provided by Demand-Responsive Transport (DRT), and modelled the travel times of DRT with analytic models, namely continuous approximation (\cite{hasif2022graph,Wang2024}). \cite{hasif2022graph} built a detailed dynamic graph representing the schedule of the conventional PT, based on real-world General Transit Feed System (GTFS), and added new ``virtual links'' with travel and waiting times obtained via Continuous Approximation modelling. \cite{Wang2024} did not adopt a detailed GTFS-based representation and resorted to a frequency-based graph: the core of this graph was a simplified representation of conventional PT, on top of which ``DRT arcs'' were added in the different regions in which DRT was assumed to be deployed, weighted by the average waiting time and travel times, computed via Continuous Approximation. The goal of \cite{Wang2024} was to find an allocation of DRT in a metropolitan region in order to decrease the inequality in the geographic distribution of accessibility. The issue with the previous two references is that they use analytic models for computing SMS performance. More precisely, they modelled SMS with Continuous Approximation~(CA), which is a theoretical model where the DRT area is represented as a rectangle, without any topology, demand is a 2-dimensional uniform density function on such a rectangle, and neither the road topology nor the routing algorithm is represented. Average ``DRT cycle times'', to traverse the entire rectangle, are computed via simplified equations, under input parameters chosen by the modeller. It is thus evident that such CA-based approaches are not intended to measure the accessibility of an actual DRT service. They are instead aimed at providing some high-level guidance, at the strategic level, for planning DRT, e.g., to deciding in which areas it could be potentially beneficial, prior to a more in-depth investigation of those areas. In this paper, instead, our effort consists in estimating accessibility from empirical observations of an actual SMS service, rather than a simplified geometric abstraction of it.
Note that, recently, the GTFS-Flex extension has allowed describing 
%to also integrate 
SMS (\cite{craig2020gtfs}). Although our estimates could thus be fed into GTFS-Flex data, for the sake of simplicity, we use plain GTFS instead.

To summarize, to the best of our knowledge, we are the first to propose a method to compute isochrone measures of the accessibility obtained via integrating SMS with a conventional PT service. The main feature of our method is that it is fully data-driven, as it allows computing accessibility based on observed trips (either from real scenarios or simulations) via statistical methods, without needing to build complex simulation models.

%%%%%%%%%%%%%%%%%%%%%%%%%%%%%%%%%%%%%%%%%%%%%%
%%%%%%%%%%%%%%%%%%%%%%% METHOD %%%%%%%%%%%%%%%
%%%%%%%%%%%%%%%%%%%%%%%%%%%%%%%%%%%%%%%%%%%%%%
\section{Method}
\label{sec:methodology}

The idea behind our method is intuitive: we aim to model Shared Mobility Services (SMS) with the same representation of conventional PT lines, so as to allow computing accessibility on a coherent model. Conventional PT is usually modelled as a time-dependent graph, which is the implicit model underlying GTFS. In this model, each PT line is represented by a set of arcs, each arc represents the passage of a bus from one stop to another. To adapt to the conventional PT model description, we wish to represent a feeder service from an area to a certain stop,\footnote{Symmetrically, we also represent an SMS service from a stop to a surrounding area.} as a set of ``virtual PT lines'', each connecting a centroid to that stop. We wish to construct such a virtual PT line as a proxy of the performance offered by SMS, i.e., in a way such that the performance of such a virtual line is similar, on average, to the performance actually experienced by SMS users.

\subsection{Accessibility}
As in \cite{biazzo2019accessibility}, the study area is tessellated in sufficiently small hexagons,\footnote{The smaller the hexagons, the more precise is the accessibility computation, but the more the computational power needed.} whose centres $\mathbf{u}\in\mathbb{R}^2$ are called centroids and denoted by set $\pazocal{C} \subseteq   \mathbb{R}^2$. Each hexagon contains a certain set of \emph{opportunities}, e.g., jobs, places at school, people. With $O_\mathbf{u}$ we denote the number of opportunities in the hexagon around $\mathbf{u}$ and with $T(\mathbf{u},\mathbf{u}', t)$ the time it takes to arrive in $\mathbf{u}'$, when departing from $\mathbf{u}$ at time $t$.
According to the isochrone-based definition of \cite{miller2020accessibility}, accessibility is the number of opportunities that one can reach departing from $\mathbf{u}$ at time of day $t$ within time threshold $\tau$:
\begin{align}
    acc(\mathbf{u},t) \equiv \sum_{\mathbf{u}'\in\pazocal{C}(\mathbf{u},t)} O_{\mathbf{u}'}, 
    \text{ where }
    \pazocal{C}(\mathbf{u},t)=\{\mathbf{u}\in\pazocal{C} | T(\mathbf{u},\mathbf{u}', t)\le \tau\}
    \text{ is the set of centroids reachable within time }\tau.
    \label{eq:acc}
\end{align}

We consider here travel times provided by PT (combining walk, fixed lines and SMS).\footnote{We deliberately make the choice of excluding other modes from our analysis, and in particular of excluding cars. It is true that people can access opportunities by car. On the other hand, given the urgent need to reduce car dependency, we argue that transport planners should prioritise the ease of accessing opportunities \emph{via sustainable modes}, rather than (or at least in addition to) the ease of reaching opportunities \emph{via whichever modes}. Indeed, if planners aimed to improve accessibility by \emph{whichever modes}, they would inevitably plan cities for cars, as cars are often the fastest way to travel. This would obviously contradict the effort of achieving sustainable cities.}
Accessibility jointly depends on the spatial distribution of opportunities (which determines number~$O_\mathbf{u}$ of opportunities around centroids~$\mathbf{u}$) and the transport system (which determines travel times~$T(\mathbf{u},\mathbf{u}',t)$). The focus of this paper is on the transport system.
By improving PT, set~$\pazocal{C} | T(\mathbf{u},\mathbf{u}', t)\le \tau\}$ can be enlarged, which consents to reach more opportunities. In the numerical results, for simplicity, the opportunities are the number of people (residents) that can be reached. We make this choice as it is sufficient to show the formalism of our method. One could easily replace the distribution of people with the distribution of other kinds of opportunities to compute accessibility indicators with different semantics. 
\footnote{
There are several accessibility indicators available, each with their pros and cons (\cite{GeursVanWee2013,GeursVanWee2023}). The focus of this work is not in the particular indicator to choose. Our method adapts to several indicators. We could have used other definitions instead of~\eqref{eq:acc}. For instance, if one wishes to capture in the accessibility indicator the variety of opportunities that are reachable (as \cite{Vale2023}), instead of their total number, it would suffice to make a small modification in~\eqref{eq:acc}. Instead of counting the opportunities~$O_{\mathbf u'}$ in hexagon~$\mathbf u'$, we could consider set~$\pazocal O_{\mathbf u'}$ of the diverse categories of opportunities. One would then replace sum~$\sum_{\mathbf u'\in \pazocal C(\mathbf u, t)} O_{\mathbf u'}$ with the cardinality of the union~$\left| \bigcup_{\mathbf u'\in \pazocal C(\mathbf u, t)} \pazocal O_{\mathbf u'} \right|$. The rest of the steps involved in our method would not change.
}

%is possible to reach.
Observe that travel times~$T(\mathbf{u},\mathbf{u}', t)$ have been so far computed on a graph representation of the transport network. However, the novel issue with which we are faced in this paper, is that SMS are not based on any network, due to their dynamic and stochastic nature. Our effort is thus to build a graph representation of SMS, despite the absence of a network model. In the following subsections, we describe our method to do so.

\subsection{Time-Expanded Graph Model of conventional PT}
\label{time-expanded graphs for pt}
\begin{figure}
\center \includegraphics[width=100mm]{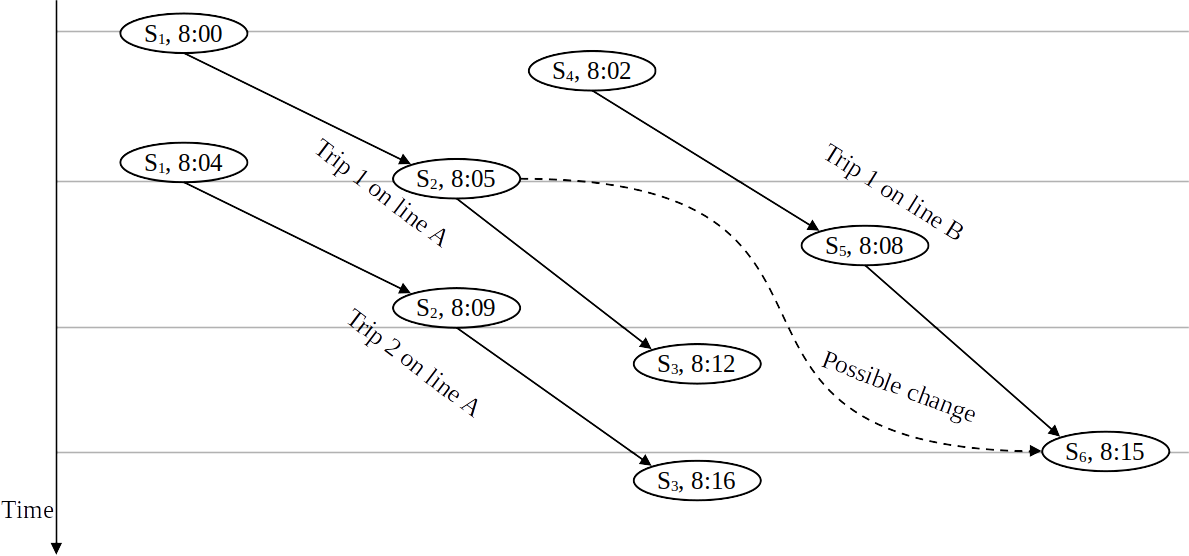}
\caption[time-expanded Graph for Dynamic Feeder Service]{Time-expanded graph, representing two trips on line A and one trip on line B, as well as a potential change.}
\label{fig:graph}
\end{figure}

We first describe the model of conventional PT. 
Inspired by~\cite{fortin2016innovative} and~\cite{hasif2022graph}, we model PT as a time-expanded graph~$\pazocal{G}$, compatible with the GTFS format. The nodes of $\pazocal{G}$ are \emph{stoptimes}. Stoptime $(\mathbf{s},t)$ indicates the arrival of a PT vehicle at a stop $\mathbf{s}\in\mathbb{R}^2$ (modelled as a point in the plane) at time $t\in\mathbb{R}$. Different trips on a certain PT line are represented as sequences of different stoptimes, as in Figure~\ref{fig:graph}. Change from a line to another is represented by a connection between stoptimes $(\mathbf{s}, t)$ and $(\mathbf{s}', t')$, which belong to the first and the second line, respectively. This ``change connection'' is added if time $T_\text{walk}(\mathbf{s}, \mathbf{s}')$ between $\mathbf{s}$ and $\mathbf{s}'$ is within a maximum tolerated walking time, e.g., 15 minutes, and if it is possible to arrive in $\mathbf{s}'$ before the departure of the corresponding vehicle, i.e., if $t+T_\text{walk}(\mathbf{s},\mathbf{s}')\le t'$. When a user departs at time $t_0$ from location $\mathbf{x}$ to location $\mathbf{x}'$, they can simply walk (but no more than the maximum walk time). Or they can walk to $\mathbf{s}$, board a PT vehicle at $t$ (corresponding to a stoptime $(\mathbf{s},t)$, use PT up to a stoptime $(\mathbf{s}',t')$ and from there walk to $\mathbf{x}'$. The arrival time
%time of arrival 
at $\mathbf{x}'$ will be $t'$ plus the time for walking.
Users are assumed to always choose the path with the earliest arrival time. Path computation is performed within CityChrone (\cite[Supplementary Information]{biazzo2019accessibility}). 
%No capacity constraints are considered.

\subsection{Integration of shared mobility into the time-expanded graph} 
\label{time-expanded graphs for shared mobility}

SMS is assumed to provide a feeder service to conventional PT. In a feeder area $\pazocal{F}(\mathbf{s})\subseteq \mathbb{R}^2$ around some selected stops $\mathbf{s}$ (which we call \emph{hubs}), SMS provide connection to and from $\mathbf{s}$. The set of centroids in such an area is $\pazocal{C}(\mathbf{s})=\pazocal{C}\cap \pazocal{F}(\mathbf{s})$.
In this section, we will focus on access trips (from a location to a PT stop) performed via SMS. The same reasoning applies to 
%can be applied on 
egress trips, \textit{mutatis mutandis}.
We assume to have a set $\pazocal D$ of observations. Each observation $i\in\pazocal D$ corresponds to an access trip and contains:
\begin{itemize}
    \item Time of day $t_i\in \mathbb{R}$ when the user requested a trip via SMS
    \label{line:information}
    \item Location $\mathbf{x}_i\in\mathbb{R}^2$ where the user is at time $t_i$
    \item Station $\mathbf{s}_i$ where the user wants to arrive via the SMS feeder service
    \item Duration $w_i$, indicating the waiting time before the user is served. In the case of dynamic Demand-Responsive Transport (DRT) (which is what we considered in the numerical results), ride-sharing,  or carpooling, $w_i$ represents the time passed between the time at which the user wishes to be picked up and the actual time of pickup. In  DRT or ride-sharing or carpooling systems where trips are booked in advance (e.g., the day before), such a time indication should be ignored. In the case of car-sharing or bike-sharing systems, such a time indication should also be ignored. In the services with no prebooking, such as the one we consider in the case study of Section~\ref{sec:results}, $w_i$ should instead be explicitly accounted, as we do.
    \item Travel time $y_i$: time spent in the SMS vehicle to arrive at $\mathbf{s}_i$.
\end{itemize}

%The intensity $\lambda(\mathbf{x},t)$ is the expected amount of arrivals per unit of time and space in an infinitesimal area and an infinitesimal time interval around $\mathbf{x}$ and $t$.

We now explain how we use such information in the case of dynamic DRT, ride-sharing and carpooling. The other cases are commented in the last paragraph of this subsection.
We interpret $y_i$ and $w_i$ as realizations of spatial-temporal random fields (\cite{Handcock1994}): for any time of day $t\in\mathbb{R}$ and physical location $\mathbf{x}\in\pazocal{F}(\mathbf{s})$, random variables $W^\mathbf{s}(\mathbf{x},t), Y^\mathbf{s}(\mathbf{x},t)$ represent the times experienced by a user appearing in $t$ and $\mathbf{x}$, willing to go to stop $\mathbf{s}$ via SMS.
Let $\hat w^\mathbf{s}(\mathbf{u},t), \hat y^\mathbf{s}(\mathbf{u},t)$ estimations of expected values $\mathbb E[W^\mathbf{s}(\mathbf{u},t)],\mathbb E[Y^\mathbf{s}(\mathbf{u},t)]$, respectively, at centroids $\mathbf{u}\in\pazocal{C}(\mathbf{s})$. We defer the computation of such estimation to the next subsection, and we now instead explain how we use such estimations to calculate ``virtual PT lines'', which we then add as additional arcs to the time-expanded graph of PT.

\begin{figure}
    \centering
    \includegraphics[width=0.35\textwidth]{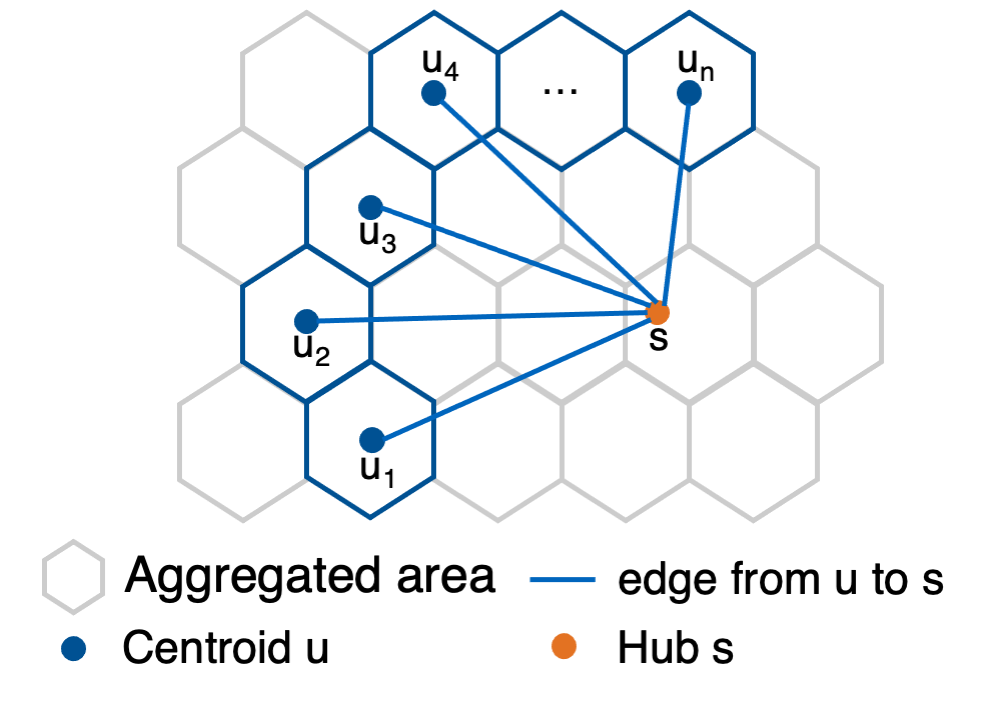}
    \caption{Hub and virtual trips provided by SMS}
    \label{fig:hub}
\end{figure}

The virtual PT line that we will use as a proxy of SMS between centroid $\mathbf{u}\in\pazocal{C}(\mathbf{s})$ and hub $\mathbf{s}$ (Figure~\ref{fig:hub}) is a sequence of ``virtual trips''. Such trips are ``virtual'' in the sense that they have not necessarily been observed in the past: they just summarize statistically the observed trips.
We make all virtual trips start at~$\mathbf{u}$ and end at~$\mathbf{s}$, and we assign to them departure times $t_j, j=1,2,\dots$ spaced by a headway~$\hat h^\mathbf{s}(\mathbf{u},t_j)$, i.e. $t_j=t_{j-1}+\hat h^\mathbf{s}(\mathbf{u},t_j)$. 
Value~$\hat h^\mathbf{s}(\mathbf{u},t_j)$ should be set so that it represents the ``latency'' a user experiences between two consecutive chances of getting the service. The average waiting time is generally assumed to be half of the latency between two consecutive services, under the assumption that users' arrival times are independent and the time between the arrival of a user and the next can be described as an exponential distribution with a certain unknown rate (\cite[(2.4.28)]{cascetta2009transportation}). This assumption is extensively used~(\cite{JaraDiaz2012waiting,Nielsen2018waiting,ZhangJie2018waiting}). \cite{Miller2020waiting} confirm that such an assumption is appropriate when the headway is relatively small, so users simply go to the stop and wait for the service to come as soon as possible (see their Section 3.3.1). They also show that such a simple behaviour is most times the most convenient for users, even better than more complicated possible user strategies. In our case study, the headways of virtual PT lines we deal with are quite small~(Figure~\ref{fig:headway_results}), and the arrival times of the requests from different users are independent of one another. Also in our case, a user simply requests a trip and wishes to be served as soon as possible. We can thus safely apply the aforementioned very common assumption and write:
\begin{align}
\label{eq:headway}
\hat h^\mathbf{s}(\mathbf{u},t_j)
=2\cdot \hat w^\mathbf{s}(\mathbf{u},t_j).
\end{align}

For each virtual SMS trip, we add a time-dependent arc to the overall graph~$\pazocal G$. We associate to that arc the trip time of the virtual SMS trip, which we set equal to~$\hat y^\mathbf{s}(\mathbf{u},t_j)$.
The origin of each arc above is stoptime $(\mathbf{u},t_j)$. Before generating stoptimes, we first fix an instant~$t_0$ (selected uniformly at random in the interval~$[00:00, 24:00]$), and then we add stoptimes before and after~$t_0$, all spaced by~$\hat h^\mathbf{s}(\mathbf{u},t_j)=2\cdot \hat w^\mathbf{s}(\mathbf{u},t_j)$:
\begin{align}
    (\mathbf{u},t_0),
    \label{eq:virtual_stoptimes0}
    \\
    (\mathbf{u},t_j)
    && \text{where }t_j= t_{j-1}+ 2\cdot \hat w^\mathbf{s}(\mathbf{u},t_{j-1})
    & \text{ for }j=1,2,& \text{ until 11:59 pm},
    \label{eq:virtual_stoptimes}
    \\
    (\mathbf{u},t_j) 
    && \text{where }t_j = t_{j+1}-2\cdot \hat w^\mathbf{s}(\mathbf{u},t_{j+1})
    & \text{ for }j=-1,-2,& \text{ until 00:00 am}.
    \label{eq:virtual_stoptimes2}
\end{align}

Correspondent stoptimes are also added to represent the arrival of the virtual SMS trips $(\mathbf{s},t_j+\hat y^\mathbf{s}(\mathbf{u},t_j))$, so that the time-expanded arcs representing virtual SMS trips connect a corresponding departure stoptime and the corresponding arrival stoptime. A similar process is applied for egress trips, \emph{mutatis mutandis}. At the end of the described process, time-expanded graph~$\pazocal{G}$ will have been enriched with stoptimes and time-expanded arcs representing SMS trips. Having done so, it is possible to reuse accessibility calculation methods for time-expanded graphs, such as CityChrone~\cite{biazzo2019accessibility}, with no modifications required. 
%In other words, the approach just described allows to decouple the estimation of wait and travel times of SMS from the calculation of accessibility. The contribution of this work focuses on the former problem while we rely
%is focused on the former problem while relies 
%on third-party tools for the latter.

\paragraph{Applications to systems with no waiting times.} What we have described so far is applicable to dynamic DRT or ride-sharing or carpooling systems (let us call them \emph{systems of type~1}). However, there are other systems in which the ``waiting time'' has no clear semantics. Let us call such other systems, \emph{systems of type 2}. We classify, as  systems of type 2, car-sharing and bike-sharing, where users pick vehicles at the dock and can thus start their trip with no waiting times. We also classify, as systems of type 2, versions of DRT, ride-sharing or carpooling services that are not dynamic, i.e., in which routes are computed offline. In this case, users book their trip in advance, e.g., the day before, and thus they do not generally suffer a relevant waiting time. The presented method should be minimally modified to account for these cases.
In systems of type 1 we consider a single time-dependent graph~$\pazocal G$ and we add the stoptimes computed with~\eqref{eq:virtual_stoptimes0}-\eqref{eq:virtual_stoptimes2} all to that graph~$\pazocal{G}$, no matter the time of day~$t$ at which accessibility~$acc(\mathbf{u},t)$ is computed (see~\eqref{eq:acc}).

We would need to do things slightly different in systems of type~2. For systems of type~2, if we want to compute accessibility~$acc(\mathbf{u},t)$ at a certain instant~$t$, we would build a graph~$\pazocal G_t$, adding simply stoptime~$(\mathbf s, t+y^\mathbf s(\mathbf u, t))$, representing the travel time to go from centroid~$\mathbf u$ to hub~$\mathbf s$, for any~$\mathbf u\in \pazocal C(\mathbf s)$ and, \emph{mutatis mutandis}, a stoptime to represent the travel time to go from the hub to the centroid in the feeder area. Observe that we do not represent waiting times in these cases, since the users of systems of type 2 do not suffer any waiting times. If we want to compute accessibility at another time~$t'$, we have to compute another graph~$\pazocal G_{t'}$.

\subsection{Estimation of Travel Times} \label{m_scarce_data}
\label{sec:kriging}

In the previous subsection, we have explained how, given estimations~$\hat w^\mathbf{s}(\mathbf{u},t)$ and~$\hat y^\mathbf{s}(\mathbf{u},t)$, we can enrich the overall graph with time-expanded arcs representing SMS.
We now explain how we construct such estimations. For simplicity, we give our explanation only for waiting times~$\hat w^\mathbf{s}(\mathbf{u},t)$ of access SMS trips only. Similar reasoning is applied to trip times~$\hat y^\mathbf{s}(\mathbf{u},t)$ and egress trips. We assume random field $W^\mathbf{s}(\mathbf{x},t)$ is approximately temporally stationary within each timeslot:
\begin{flalign}
    && W^\mathbf{s}(\mathbf{x},t)=W^\mathbf{s}(\mathbf{x},t_k),
    &&
    \forall \mathbf{x}\in\mathbb{R}^2, \forall t\in[t_k, t_{k+1}[, \forall \text{ station }\mathbf{s}
    &
    \label{eq:stationary}
\end{flalign}

For any timeslot, we thus just need to find estimation~$\hat w_{t_k}^\mathbf{s}(\mathbf{x})$ of the expected values of random field~$W_{t_k}^\mathbf{s}(\mathbf{x}) \equiv W^\mathbf{s}(\mathbf{x},t_k)$.
First, we collect the observations $\pazocal D$ that fall onto time-slot $[t_k, t_{k+1}]$:
\begin{equation}
        \pazocal D_{t_k}^\mathbf{s}
        \ \ \ \equiv
        \left\{
            \text{observation }i=(\mathbf{x}_i,w_i,y_i) | i\in\pazocal D, t\in[t_k,t_{k+1}[,
            i \text{ is related to an access trip to }\mathbf{s}
        \right\}
\end{equation}
Estimation $\hat w_{t_k}^\mathbf{s}(\mathbf{x})$ is computed by Ordinary Kriging (\cite{Srivastava1989,Geosciences}) on observations $\pazocal D_{t_k}^\mathbf{s}$ as a convex combination of observations $w_i$:
\begin{equation}
    \hat w_{t_k}^\mathbf{s}(\mathbf{x}) = \sum_{i\in\pazocal D_{t_k}^\mathbf{s}} \lambda_i\cdot w_i
    \label{eq:kriging-estimation}
\end{equation}
    
In short (details can be found in \cite[Sec.~19.4]{Chiles2018ok}), coefficients $\lambda_i$ are computed based on a \emph{theoretical semivariogram} function $\gamma_{t_k}^\mathbf{s}(d):\mathbb R^+ \rightarrow \mathbb R^+$, which is obtained as a linear regression model, with predictors $d_{i,j}$ (distances between all pair of observations) and labels $\gamma_{i,j}$, which are called \emph{experimental semivariances} (\cite[Ch.~5]{olea2000geostatistics}):
\begin{equation}
    \gamma_{i,j} \equiv \frac{1}{2}\cdot (w_i-w_j)^2
    \label{eq:gamma_ij}
\end{equation}
    
 The underlying assumption here is that the correlation between waiting times in two different locations vanishes with the distance between such locations.\footnote{Note that we talk here about the correlation between the waiting time of observation~$i$ and the waiting time of observation~$j$ and we are assuming that such a correlation vanishes with their mutual distance~$d_{i,j}$. This is not to be confused with the dependence between the waiting time of observations and the distance to the hub, which may not exist, as later shown in Figure~\ref{fig:corr_pairplot_wt}} The theoretical semivariogram gives the ``shape'' of this vanishing slope. The procedure to derive the theoretical semivariogram and the coefficients~$\lambda_i$ are explained in Appendix~\ref{sec:derivation}. Formula~\eqref{eq:kriging-estimation} implicitly implies that different observations~$w_i$ contributes differently to the estimation of waiting time in location~$\mathbf{x}$, and the contribution of observations closer to~$\mathbf{x}$ is given higher weight~$\lambda_i$. Under hypothesis on spatial stationarity and uniformity in all directions (\cite{Columbia}), Theorem~2.3 of \cite{Szidarovsky1985}\label{line:unbiased} proves that Kriging is an asymptotically biased estimator: as the number of observations goes to infinite, $\hat w_{t_k}^\mathbf{s}(\mathbf{x})$ tends to the ``true'' $\mathbb{E}[W_{t_k}^\mathbf{s}(\mathbf{x})]$.

Note that, by means of interpolation on a limited set of observed trips, the method described here allows inferring the potential to access opportunities, also via trips that may not have been observed yet.

%%%%%%%%%%%%%%%%%%%%%%%%%%%%%%%%%%%%%%%%%%%%%%%%%%%%%%%%%%%%%%%%%
%%%%%%%%%%%%%%%%%% IMPLEMENTATION %%%%%%%%%%%%%%%%%%%%%%%%%%%%%%%
%%%%%%%%%%%%%%%%%%%%%%%%%%%%%%%%%%%%%%%%%%%%%%%%%%%%%%%%%%%%%%%%%
\section{Implementation}
\label{sec:implementation}
\begin{figure}
    \noindent\hspace{0.5mm}\includegraphics[width=149mm]{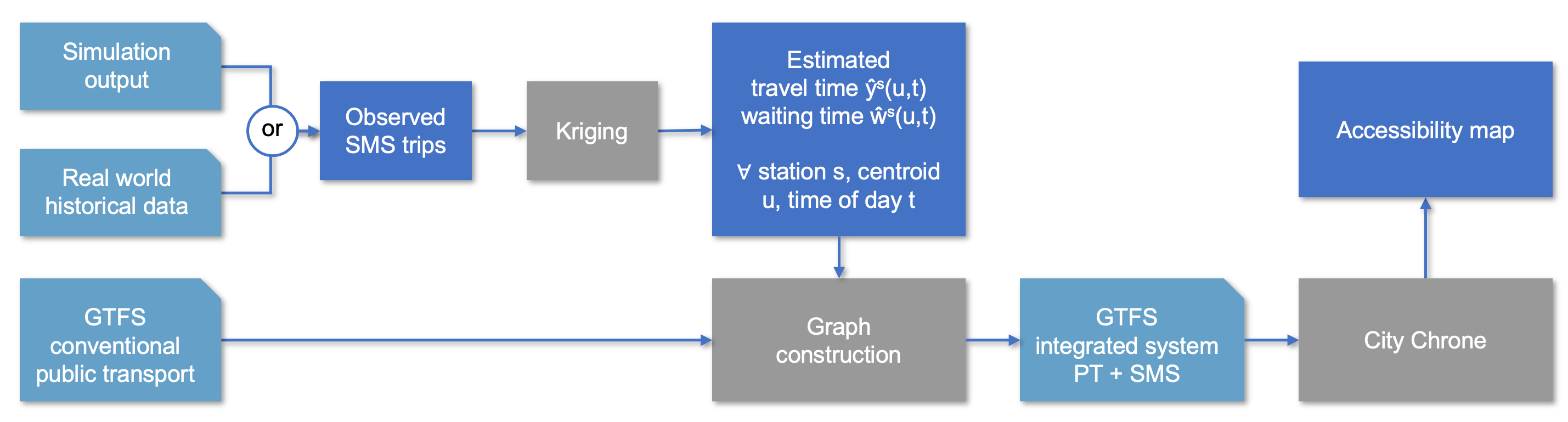}
    \caption[Schema of Implementation for SMS Services]{Implementation pipeline. Note that input data can come either from simulation or a real scenario. CityChrone is the tool developed by \cite{biazzo2019accessibility} and that we use in our work. However, other tools, such as Conveyal Analysis and Remix, could be used instead.}
    \label{fig:workflow}
\end{figure}

%\textbf{DRT2GTFS Pipeline}
    The method of Section~\ref{sec:methodology} is implemented in a Python pipeline, which we release as open source (\cite{githubrepo}) and is depicted in Figure~\ref{fig:workflow}. 
    \begin{enumerate}
        \item We first get centroids and cells performing the tessellation via CityChrone software~\cite[Figure~1]{biazzo2019accessibility}.
        \item We read the file containing the observations (SMS trips). Such a file can be a simulation output or measurements of real SMS. Each observation includes the information listed in Sec.~\ref{time-expanded graphs for shared mobility}. Observations are stored in a dataframe.
        \item We assume SMS is deployed as feeder (as it is the case for the MATSim simulation on which we perform our analysis). Therefore, we can classify every SMS trip as either access or egress, depending on whether the origin or the destination is a conventional PT stop.
        \item To establish feeder area $\pazocal{F}(\mathbf{s})$ around any stop~$\mathbf{s}$, we find among the observations $\pazocal D$ the furthest cell from $\mathbf{s}$ in which a trip to/from $\mathbf{s}$ has occurred.\footnote{
    In the presence of outliers (which we do not have in our numerical results), some criterion should be established to eliminate them. Such criterion could be, for example, a maximum distance threshold that can never be exceeded. Another criterion would instead be to use some knowledge about the SMS configuration. For instance, if areas~$\pazocal F(s)$ are established during the planning of SMS, one could reuse such information.}
         All cells within such a distance are assumed to be in $\pazocal{F}(\mathbf{s})$. Observe that feeder areas of different hubs may overlap.
        \item We group observations in timeslots (Figure~\ref{fig:distribution_wt_access}).
        \item In each time slot $[t_k,t_{k+1}[$ and each centroid $\mathbf{u}$ in $\pazocal F(\mathbf{s})$, we perform Kriging (Sec.~\ref{m_scarce_data}) via library $\texttt{pyInterpolate}$ (\cite{molinski2022pyinterpolate}) to obtain estimations $\hat w^\mathbf{s}(\mathbf{u})$ and $\hat y_{t_k}^\mathbf{s}(\mathbf{u})$.
        \item We create stoptimes and time-dependent arcs using the estimations above, as specified in ~\eqref{eq:virtual_stoptimes}. We add stoptimes and time-dependent arcs to the GTFS data of conventional PT, following the specifications in \cite{gtfsReference}.
        \item We give the obtained GTFS file as input to CityChrone, which will give us accessibility scores in all the centroids.
    \end{enumerate}

%%%%%%%%%%%%%%%%%%%%%%%%%%%%%%%%%%%%%%%%%%%%%%%%%%%%%%%%%%%%
%%%%%%%%%%%%%%%%%%%%% RESULTS %%%%%%%%%%%%%%%%%%%%%%%%%%%%%%
%%%%%%%%%%%%%%%%%%%%%%%%%%%%%%%%%%%%%%%%%%%%%%%%%%%%%%%%%%%%
\section{Exemplary case study}
\label{sec:results}

\begin{table}
   
  \begin{center}
    \caption{Parameters used for the numerical results.}
     \vspace{2ex}
    \label{tab:parameters}
    \begin{tabular}{l|c|l} % <-- Alignments: 1st column left, 2nd middle and 3rd right, with vertical lines in between
      \textbf{Parameter} & \textbf{Value} & \textbf{Reference}\\
      \hline
      Side of a hexagon (tessellation) & 1 km & \cite{badeanlou2022ptanalysistool}\\
      $\tau$ (Equation~\eqref{eq:acc}) & 1 hour & \cite{badeanlou2022ptanalysistool}\\
      Total number of SMS trips & 14700 & 
      \cite{Chouaki2023}
      \\
        - as access towards PT & 5289 & 
        \cite{Chouaki2023}
        \\
        - as egress from PT & 9412 & 
        \cite{Chouaki2023}
        \\
    Maximum tolerated walk time & 15 minutes &\\
    Total number of hubs & 16 &\\
    Walk times & & Computed via OpenStreetMap
    \\
    Travellers in the considered area & 76k travellers & From the simulation of \cite{Chouaki2023}
    \\
    Average request rate served by SMS in the entire region & 13.6 requests / minute 
    & \cite{Chouaki2023}
    \\
    SMS fleet size \begin{small}(calculated in order to minimize rejections)
    \end{small} & 1600 & \cite{Chouaki2023}
    \\
    Headway of added rail and tramway lines 
    &
    2.4 minutes
    &
    \cite{Chouaki2023} and \cite{GPE}
    % If the information on the Km trvaled by SMS is not available, we can remove it, it is not a big deal.
    %\\
    %Kilometers traveled by the SMS fleet & 
    %XX
    %& \cite{Chouaki2023}
    \end{tabular}
  \end{center}
\end{table}

\subsection{Data Source of the observations} 
\label{Scenariodescribtion}

\begin{figure}  
\center    
\includegraphics[width=0.65\textwidth]{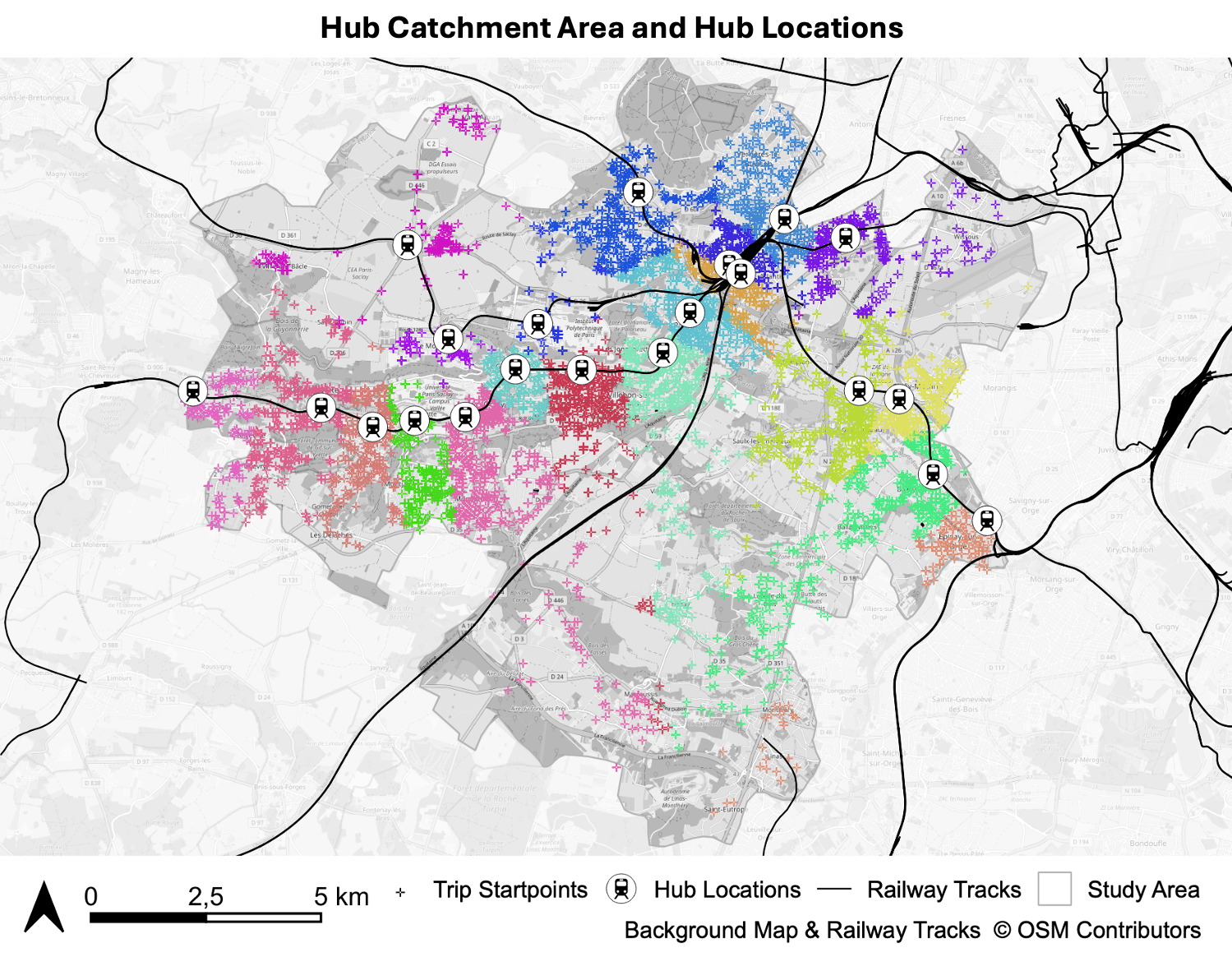}
\caption[Hub Catchment Areas and Hub Locations]{Each dot corresponds to the origin of one trip observed in the simulation. The differentiation in the colour of the dots associates them to the hub at which the user transferred from SMS to PT. Solid black lines represent all tracks, i.e., metro/train lines, but also side tracks and rail yards. We only depict the metro/train hubs considered in this study, in which transfers between SMS and PT are possible. Observe that, in some cases, SMS can pick up passengers far from the hub in which they will be dropped off.}
\label{map:drt_hubs}
\end{figure}

The dataset of observed trips used in this study was obtained from a MATSim simulation. Instead of making up a new scenario specifically for this paper, we preferred to apply the presented method to results obtained from a separate piece of work (\cite{Chouaki2023}), where a dedicated effort was focused on obtaining a solid and accurate simulation scenario.
Detailed analyses on the mobility patterns and mode shares in the area are documented in \cite{Chouaki2023}. The simulation includes the future rail infrastructure, consisting of new automated subway lines and tramway lines, which will be deployed within the Grand-Paris Express project, in the Paris-Saclay area, located in the south of Paris in the Île-de-France region, depicted in Figure~\ref{map:drt_hubs}.
\rev{}{}{The figure allows appreciating that zones associated to the different hubs are heterogeneous in terms of size, as well as in terms of density (number of origin locations in the unit of space).}\footnote{
\rev{}{}{Note that an alternative visualization could have been a density map, but that would have not allowed one to appreciate such a heterogeneity, which is instead clear with the colored dots in the figure. Indeed, in a heatmap, the information about the localization of the different origins would be lost.}
}
The headways of the new lines, reported in Table~\ref{tab:parameters}, are based on prospective information from  Saclay's transport planning agency (\cite{GPE}). The information about their variation along the day was not available. The lines that exist today are instead built following current GTFS data. An SMS service, namely Demand-Responsive Transport, is simulated, operating as a feeder for rail-based public transport. 
%The motivation behind this operational scheme is that under the absence of data regarding how bus lines will adapt to new rail ones, the presence of an on-demand SMS would allow travellers to access new stations and thus better estimate the attractiveness of new rail lines. 
The SMS service serves door-to-rail-stop and rail-stop-to-door trips in the area of Paris-Saclay.

This simulation is obtained starting from a synthetic population of the Île-de-France region, generated exclusively from open data, and a MATSim simulation for this population (\cite{horl_synthetic_2021}). Four future subway lines planned for the region in the scope Grand-Paris Express project are added to this simulation, as well as the tramway line $T12$, which was under construction in the Paris-Saclay area at the time of the study. After these modifications on the PT offer, a region wide simulation is performed to identify a preliminary impact of these lines on traveller decisions. Afterwards, for the simulation of the SMS service, a focus is performed on the Paris-Saclay area by cutting the road and PT network around it and leaving only agents that travel in this area. The SMS service is simulated in competition with the other modes (walk, car, bike, PT). This means that the identified trips that are used later on reflect the attractiveness of the service.

In order to appropriately size the service, a first set of simulations is performed by setting a maximum waiting time of 10 minutes and varying the fleet size. In this first set of simulations, trip requests that cannot be served with a waiting time less than the fixed threshold are rejected. A fleet size of 1,600 vehicles was determined as the minimum necessary to ensure less than 5\% rejection rate. However, considering rejected requests in the accessibility calculation method presented below is difficult.
%
%\footnote{This limitation is not related to our method. It is rather related to the way in which PT is modelled into the common GTFS format and to the way in which travel times are computed on top of it. Indeed, when planners compute shortest paths on the time-dependent graph representing the GTFS schedules (see Section~\ref{time-expanded graphs for pt}), the capacity of vehicles is not considered.}
%
%
Hence, an additional simulation with 1,600 vehicles is performed, in which rejections are disabled, i.e., the service is forced to accept all requests. This is the simulation from which we collect that data that are analysed in this paper. This simulation set-up causes in some case very high waiting times for requests that cannot be integrated efficiently into the vehicle schedules. Such high waiting times impact the accessibility computation, and tend to decrease the accessibility scores. This is desirable, as it is correct to penalize the accessibility measure to account for request that are not easily served.

The SMS simulations used here should be considered as an example of input data to feed our accessibility calculation method. The on-demand service scenario itself has seen various improvements, for instance, by performing a full cost-benefit analysis and using more efficient dispatching algorithms (\cite{carreyre2024demand}). The fleet size could be optimized for achieving cost efficiency. Further refinements in realism can be achieved in the future, for instance, by making use of driver shifts (\cite{zwick_shifts_2022}) instead of assuming a service running throughout the whole day. 
The active fleet could be modulated over the day to adapt to the level of demands and SMS could be deactivated when demand is extremely low, e.g., during the night. All these improvements are related to the SMS plan and the simulation, rather than the method proposed in this paper.
Note that it is not possible to know if the information we received from the planning authority (\cite{GPE}) will match with the actual service that will be deployed. It will thus be interesting to run again the same analyses of this paper, based on updated information and, ideally, based on real observed SMS trips.

We consider people as opportunities, i.e., parameter $O_{\mathbf u'}$ in~\eqref{eq:acc} denotes the number of residents in hexagon~$\mathbf u'$. In other words, we compute the accessibility of a location as the number of people that can be reached in a limited time. This type of accessibility indicator is called \emph{\textbf{Sociality Score}} in the literature (\cite{biazzo2019accessibility}) and is largely adopted as a ``proxy for the available number of opportunities, adopting the assumption that the concentration of people is the necessary precondition for any development of amenities'' (\cite{Bossauw2019}).

Scenario parameters are shown in Table~\ref{tab:parameters}. While absolute results may change when changing parameters such as the side of the hexagon and the time threshold~$\tau$, the method (which is the focus of this paper) would remain unchanged. Time threshold~$\tau=1h$ was chosen because it is in the same range of the commuting times in the Paris region. Larger values would lead to overoptimistic evaluations of accessibility, since they would count opportunities that in reality are excessively time-costly to reach. Smaller values would also be meaningful, and would lead to a more conservative assessment of accessibility. However, playing with different values of~$\tau$ would obviously change the results, but not the method, and is thus outside the scope. The choice of the side of a hexagon is subject to a trade-off: the smaller, the finer the analysis of the accessibility but, at the same time, the larger the number of total hexagons, and thus the larger the computation time. We chose a side of 1 kilometre, which is a good compromise, since it allows capturing the changes in accessibility over the space, while keeping computation time reasonable.

\subsection{Estimation of Waiting and Travel Times}
\label{sec:quality}

\begin{figure}      
\center\includegraphics[width=120mm]{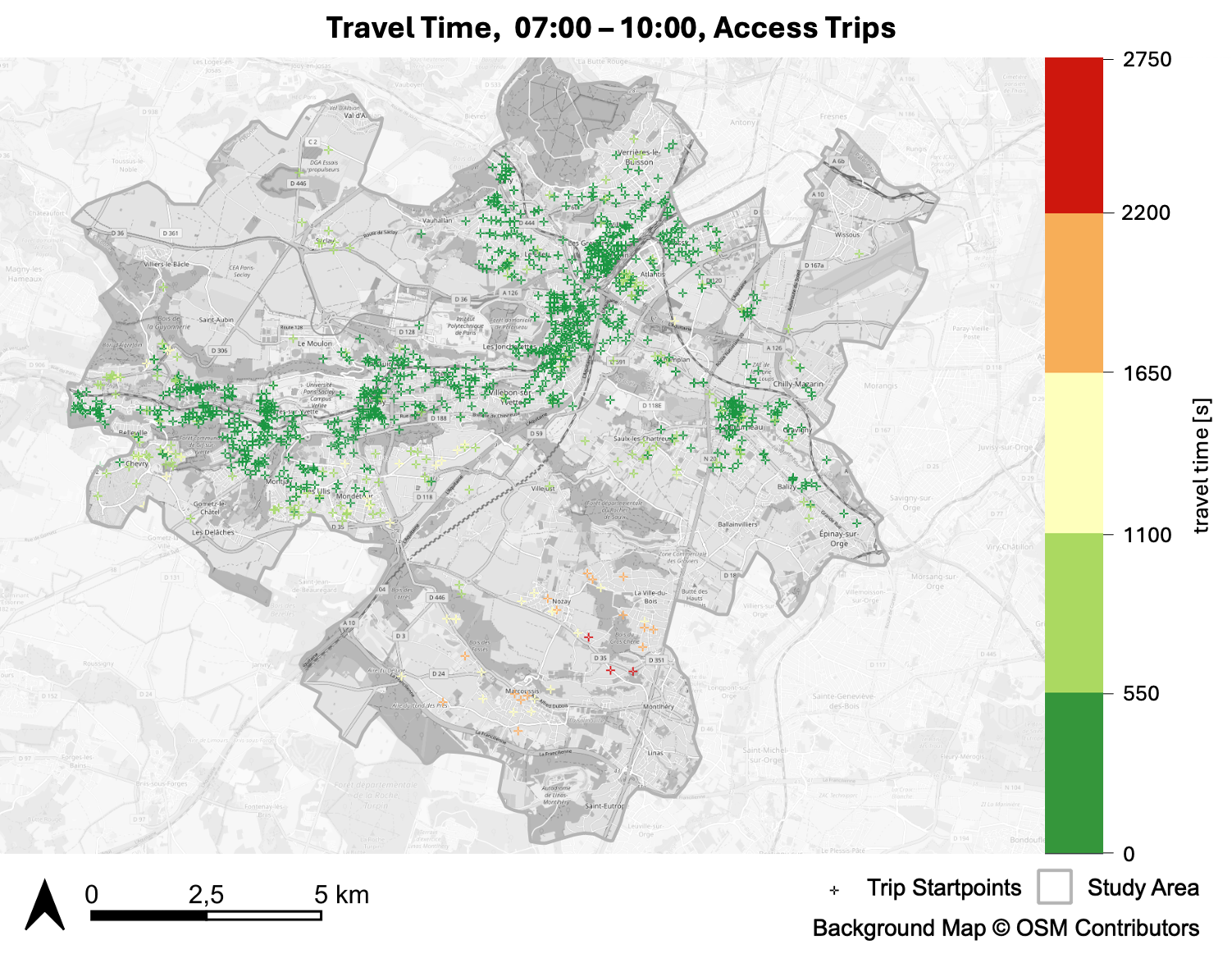}
\caption[Spatial Trend Travel Time]{Spatial Trend of the Travel Time for trips access PT in the interval [7:00 - 10:00] (evening and off-peak show similar trends). }
\label{map:tt_space}
\end{figure}

\begin{figure}      
\center\includegraphics[width=100mm]{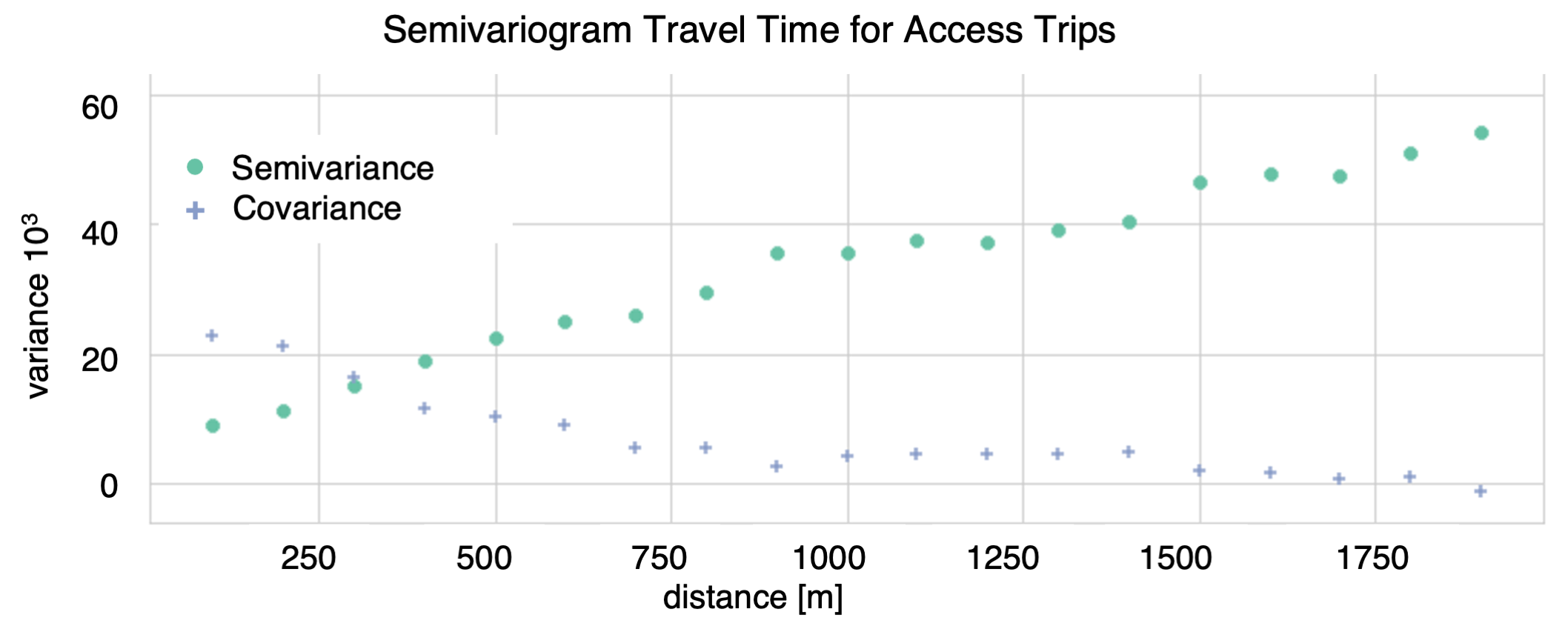}
\caption[Semivariogram Travel Time of Access Trip Observations]{Spatial correlation of travel time observations}
\label{semivar:tt}
\end{figure}

\begin{figure}      
\center
\includegraphics[width=128mm]{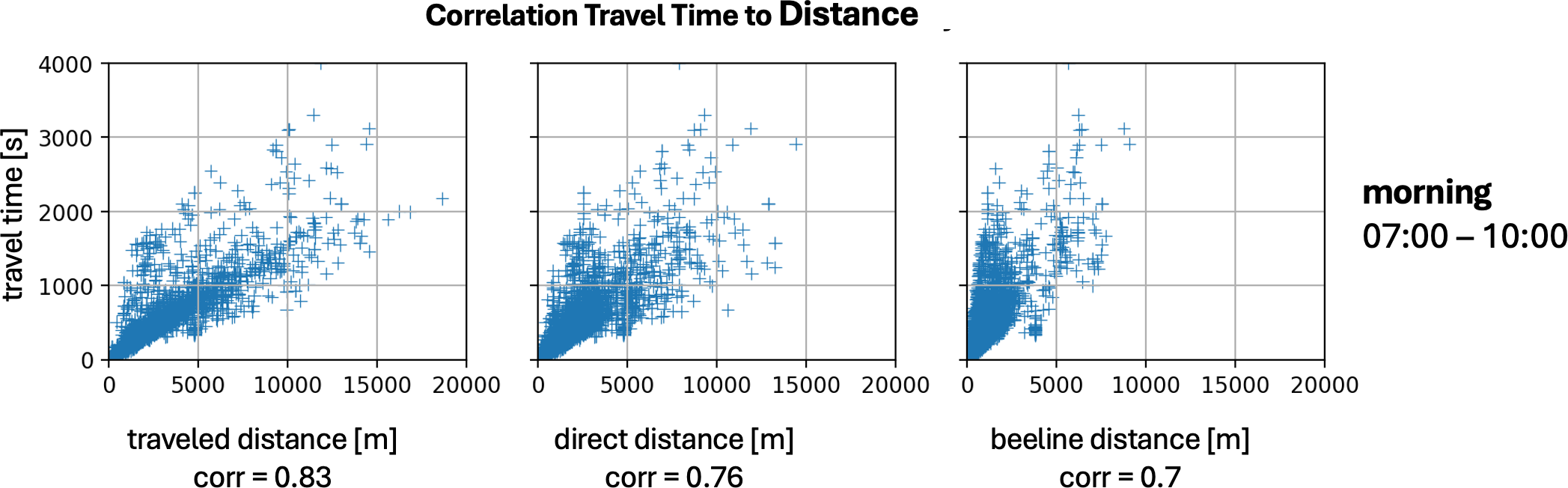}
\caption[Pair-plot Travel Time - Travelled Distance, Direct Distance, Beeline Distance]{Relation between travel time and distance measures. Travelled distance is the actual Km travelled by the user inside the SMS vehicle. Direct distance is the one from the shortest road network road from the origin centroid to the hub. The beeline is the Euclidean distance. The plots are similar over the other periods of the day and are omitted in the interest of space.}
\label{fig:corr_pairplot_tt}
\end{figure}
 
Figure~\ref{map:tt_space} clearly shows that there is spatial correlation between observations, i.e., the closer the observations, the similar the values of observed travel times, which is favourable to the use of Kriging.
The high spatial autocorrelation is induced by the high correlation between travel times and travelled distance (Figure~\ref{fig:corr_pairplot_tt}).
 A complementary way of looking at correlations is the semivariogram (Figure~\ref{semivar:tt}). A detailed derivation of the semivariogram is explained in Appendix~\ref{sec:derivation}, but already intuitively, the figure shows that, if we take any pair of observations, the higher the distance among them, the higher the semivariogram, which implies that two observations are less and less correlated, the further away they are from each other. This is also a favourable condition to the application of Kriging. 

The following figures concern SMS trips toward/from all hubs, without distinguishing between hubs.
Figure~\ref{fig:corr_pairplot_tt} contains a negative result: travel times (figures on the right) do not appear to be spatially stationary (the distribution of values measured close to the related PT stops is different from further). Therefore, our estimations are not guaranteed to be asymptotically unbiased (Sec.~\ref{sec:kriging}). We report this precision about asymptotic behaviour for the sake of mathematical coherence. However, this does not mean that our computation is ``wrong''. Indeed, as explained by~\cite[bullet 3 of p173]{oliver2007geostatistics}, perfect spatial stationarity is very often violated in practical conditions, but Kriging is still employed in such cases. Indeed, a nice feature of Kriging is that the estimation of a quantity at a point (such a quantity is travel time or waiting time in our case) is a weighted sum of the samples of that quantity at close points, and the closest points have the largest weight~$\lambda_i$ (see~\eqref{eq:kriging-estimation}). Therefore, it is tolerated to make large errors for points that are far away, since their impact on the estimation is quite limited. Additionally, we can limit the set of samples to include into the estimation~\eqref{eq:kriging-estimation} to only those within a maximum distance, called \emph{range}. This is what we do setting a range of~$3000m$ (see Appendix~\ref{sec:derivation}). Perfect asymptotic unbiasedness thus remains a purely theoretical property and models can be good in practice, even if they do not exhibit it \cite[\S{}5.1]{gow2024empirical}. In our future work, we will explore indirect estimation of travel times through other indicators, e.g., the detour factor of SMS, seeking those that may respect the requirements for the unbiasedness of Kriging.

In the Appendix (Figures~\ref{fig:corr_pairplot_wt}-\ref{semivar:wt}), we will observe that, due to the weaker correlation between waiting times and physical travelled distance, the spatial autocorrelation is weaker, compared to the case of travel times. 
Such an autocorrelation is less strong for waiting times. In any case, also in the case of waiting times, similarity diminishes with the distance, which is favourable for the use of Kriging.

\subsection{Improvement of Accessibility Brought by Shared Mobility Services}
\label{sec:improvement}

\begin{figure}
\center\includegraphics[width=148mm]{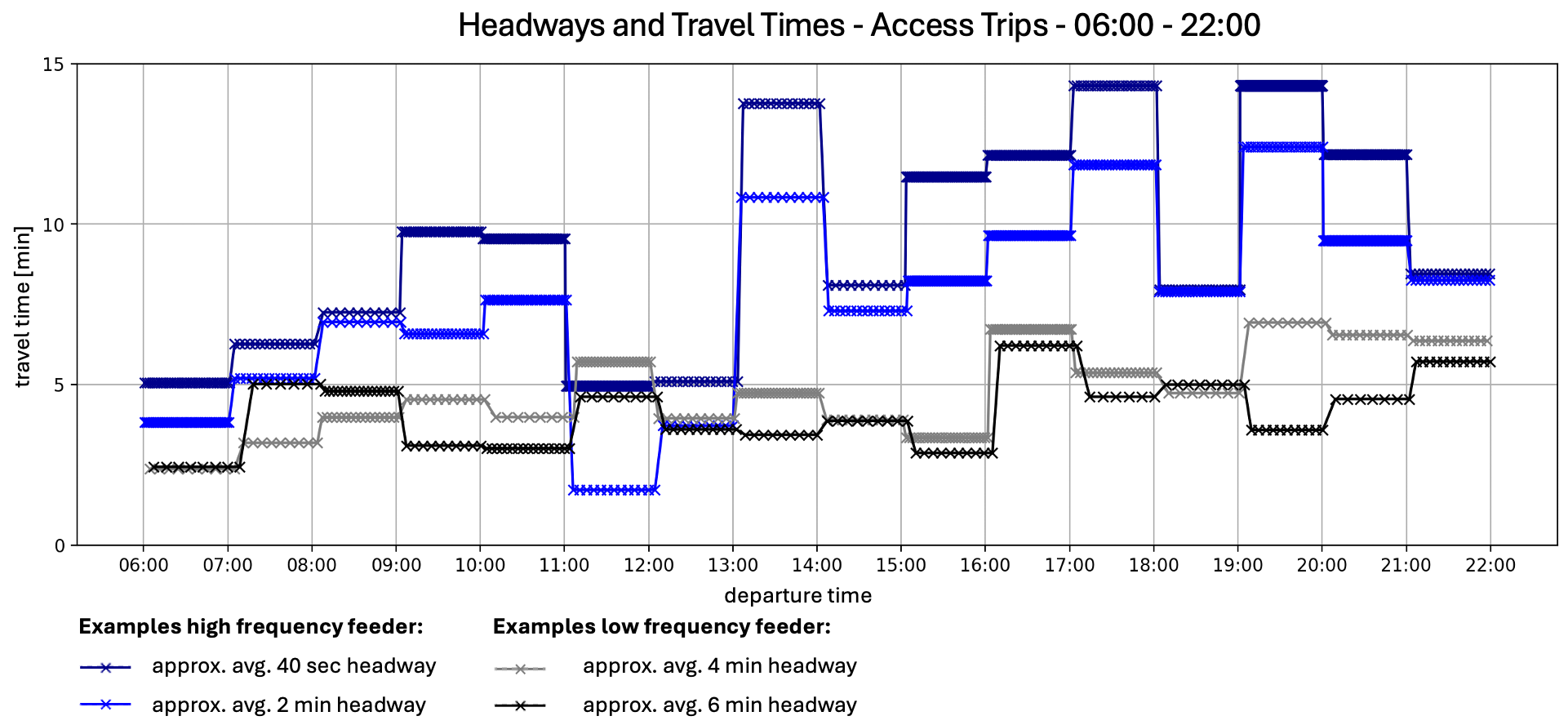}
\caption[Headway and Travel Times in output GTFS]{Headway and Travel Times of four examples of virtual SMS trips. Each departure time of a virtual SMS trip is indicated by a cross. Therefore, the distance between cross is the estimated headway. The respective travel time is indicated by the y-axis value.}
\label{fig:headway_results}
\end{figure}

Figure~\ref{fig:headway_results} depicts headway and travel times of the virtual SMS trips added to PT graph~$\pazocal G$. The figure shows that the headway and the travel times, estimated as in Section~\ref{time-expanded graphs for shared mobility}, capture the dynamics of SMS performance over the day.
Each cross represents the departure of a virtual SMS trip, from a centroid~$\mathbf u$ to a hub~$\mathbf s$. Virtual SMS trips of four different pairs (centroid, hub) are represented. The trips of each of the four pairs correspond to a line in the plot. The time between a cross and the next represents the headway~$\hat h^\mathbf s(\mathbf u,t_j)$, computed as in~\eqref{eq:headway}. In other words, the crosses are positioned in each of the time instants computed in~\eqref{eq:virtual_stoptimes0}-\eqref{eq:virtual_stoptimes2}. The values in the vertical axis represent the travel time associated to each virtual SMS trip, i.e., value~$\hat y^\mathbf s(\mathbf u, t_j)$, as computed in Section~\ref{time-expanded graphs for shared mobility}. Since such travel times are estimated per each 1-hour timeslot (see Section~\ref{time-expanded graphs for shared mobility}), the value of~$\hat y^\mathbf s(\mathbf u, t_j)$ varies at each hour, which explains the jumps in the plot.\footnote{\rev{}{}{Observe that the discontinuity in the estimated travel times, visualized in the jumps of Figure~\ref{fig:headway_results}, is an artifact of our method. Such a discontinuity would be reduced if we adopted a smaller length of the timeslot in which we perform the travel time estimation. However, in that case, we would have less observations per time slot, which would reduce the quality of the estimation. Our method thus requires the analyst to solve this trade-off. This can be considered a limitation of our method, which could be overcome in future work via some intelligent smoothing operation, or applying Kriging in the space-time, instead of applying Kriging in space only, at each timeslot. In our case, we chose 1h timeslot as it is sufficiently small to account for the temporal evolution of travel time, while also ensuring enough observations per timeslot, to enable stable estimations.}}

We compute accessibility on time-expanded graph~$\pazocal G$, obtained by adding, to the time-expanded graph representing the schedule of conventional PT, the virtual SMS arcs with the timings depicted in Figure~\ref{fig:headway_results}. Note that accessibility varies with the time of day~\eqref{eq:acc}. However, in the following figures, we show averages over the entire day.

\begin{figure}
\center
\includegraphics[width=0.9\textwidth]{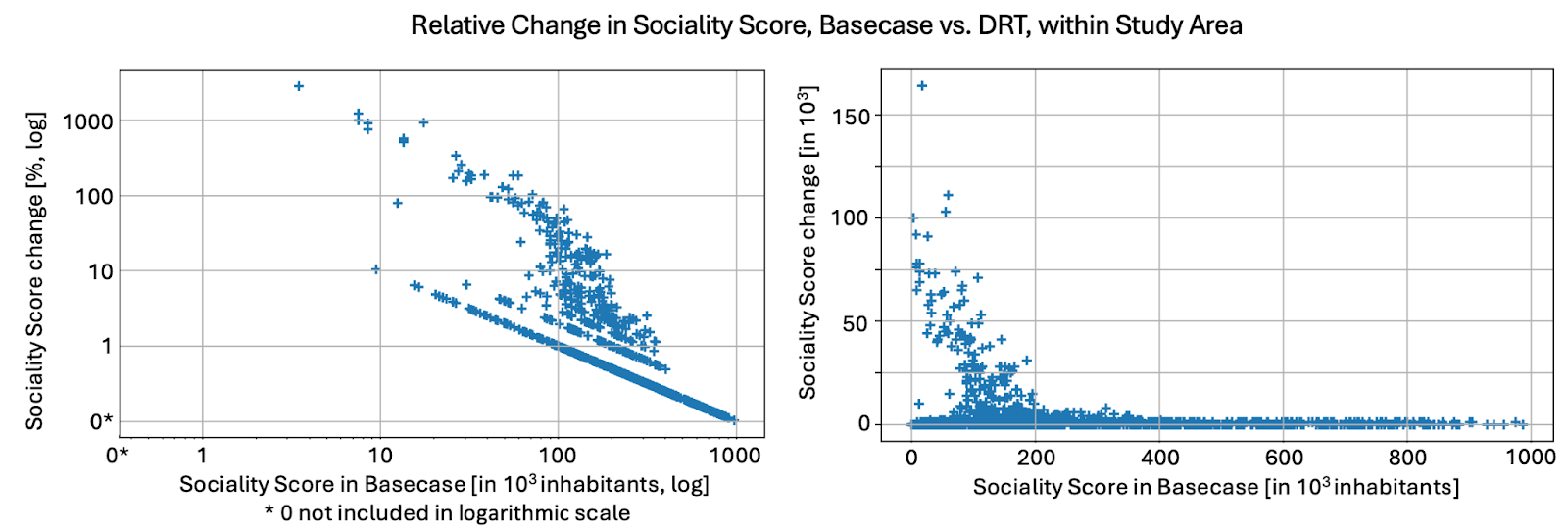}
\caption{ Improvement of the sociality score 5:00 - 23:00. Left: relative improvement (compared to the Baseline), axes are in logarithmic scale. Right: absolute improvement  (compared to the Baseline), axes are in linear scale. }
\label{fig:improvement}
\end{figure}

The improvement of PT accessibility (in terms of sociality score) achieved via the introduction of SMS is evident, as shown in Figure~\ref{fig:improvement}. Each point of the plot corresponds to a hexagon, where the x-value is the accessibility of the Basecase (conventional PT only) and the y-value is the accessibility achieved after integrating SMS. Accessibility improvement goes up to 100 thousands additional reachable people per hour (Figure~\ref{fig:improvement}-right). Improvement is concentrated in particular in the hexagons that were suffering from low accessibility in the Basecase. The average relative improvement is 18.5\% for the hexagons that, in the Basecase, could reach less than $200$ thousand people per hour. For the hexagons suffering from extremely low accessibility in the Basecase, the improvement raises for some hexagons between 100\% and 1000\% (Figure~\ref{fig:improvement}-left).
This suggests that deploying SMS is a promising way to reduce the geographical inequality of the accessibility distribution, confirming the theoretical findings of \cite{Wang2024}.

\begin{figure}
\center
\includegraphics[width=0.72\textwidth]{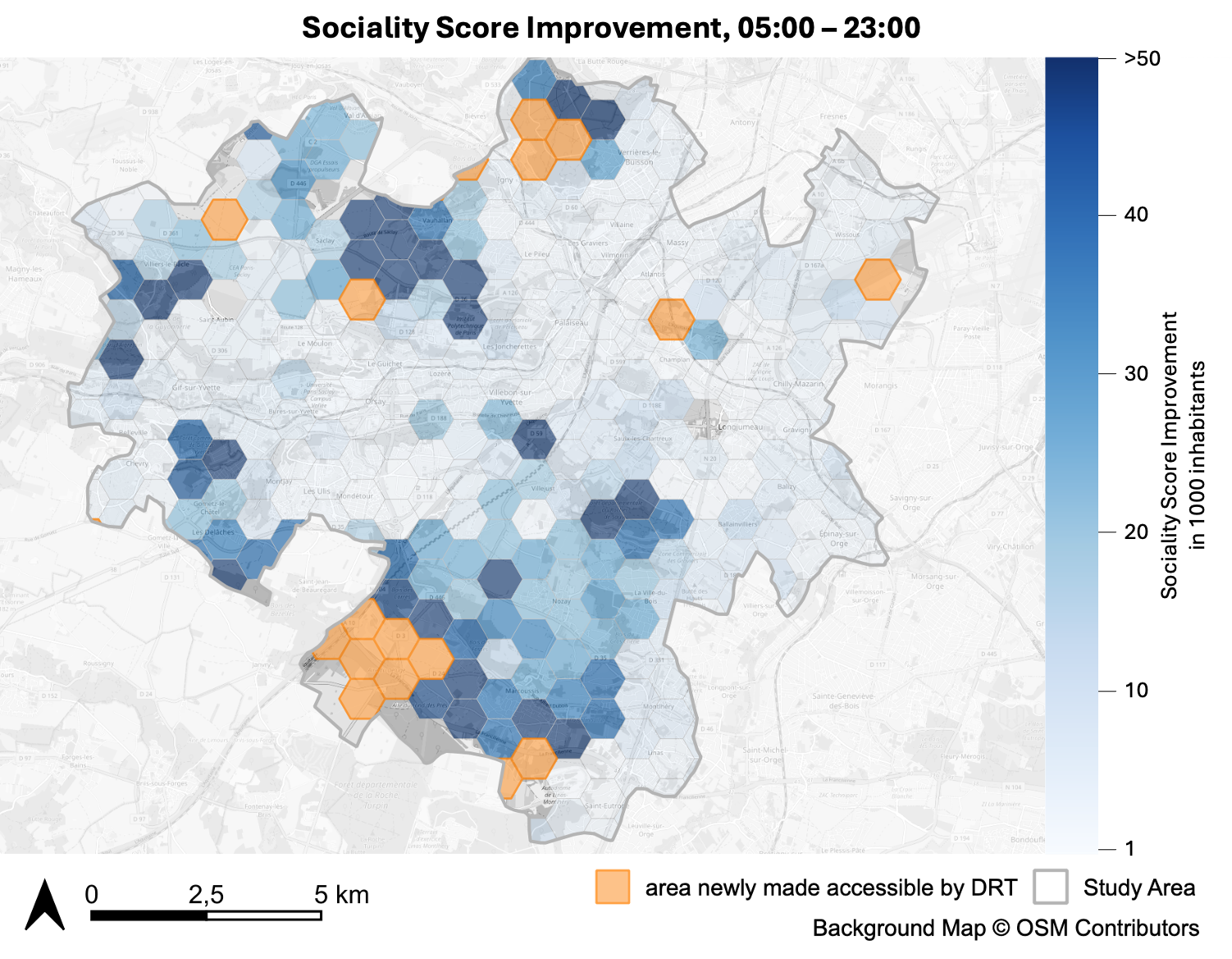}
\caption{Sociality Score Improvement- Full Day 05:00–23:00}
\label{map:sc_diff_bc_drt_fd}
\end{figure}

Figure~\ref{map:sc_diff_bc_drt_fd} visualizes the geographical distribution of the accessibility improvement. Darker colours indicate higher improvement, in terms of additional reachable people per hour, thanks to integrating SMS with PT. Observe that some hexagons (the orange ones) did not have easy access to PT, i.e., inhabitants of those hexagons needed to walk more than 15 minutes to reach a stop in the Basecase. After integrating SMS, they are instead connected to PT. Accessibility improvement is concentrated in some subregions. By comparing the map in Figure~\ref{map:sc_diff_bc_drt_fd} with the accessibility of the Basecase (that we omit in the interest of space), we observe that accessibility improvement is distributed in the areas with low accessibility in the Basecase.

%%%%%%%%%%%%%%%%%%%%%%%%%%%%%%%%%%%%%%%%%%%%%%%%%%%
%%%%%%%%%%%%%%%%% DISCUSSION %%%%%%%%%%%%%%%%%%%%%%
%%%%%%%%%%%%%%%%%%%%%%%%%%%%%%%%%%%%%%%%%%%%%%%%%%%
\section{Discussion and perspectives}
\label{sec:discussion_and_perspectives}
This paper advances the state of the art in the transport domain by proposing, to our knowledge, the first method to compute the accessibility resulting from integrating SMS and PT. Thanks to the proposed method, transport authorities, planners or operators can quantify how many additional opportunities people can reach in a certain time, thanks to the introduction of SMS. Being able to perform such quantification is extremely important to communicate the benefits of SMS in a way that is immediately understood by the public. This can help justify the cost that authorities might decide to allocate to SMS.

The present work focuses on describing the methodological contribution and does not pretend to cover all aspects of accessibility and SMS. Indeed, our work has the limitations that we will discuss below, and at the same time it opens numerous research directions, which would be impossible to exhaust in one paper and will be pursued in the future work.

It is worth stressing that the proposed method is \emph{data-driven}, since accessibility is computed via statistical analysis on previously observed SMS trips. On the one hand, a consequent advantage is that our accessibility indicator implicitly captures the characteristics of the SMS service and their impact on accessibility, without requiring dedicated effort from the modeller. For example, if the fleet is under-dimensioned, this will be reflected in higher measured waiting times (since vehicles, which are capacitated, need to make longer detours before picking up each user). This will automatically be reflected in a reduction of the accessibility indicator. On the other hand, the data-driven nature of our method could be a disadvantage, since it is not straightforward to analyse the causes behind certain dynamics of accessibility. For instance, suppose that our method indicates that, at a certain time of the day, the accessibility drops in a certain area. Such an observation would just be a mechanical consequence of degraded travel times contained in the dataset of previously observed trips. Our indicator could not give any useful ``diagnostic'' information, to understand the reason of the accessibility drop (Congestion on the road? Sudden increase in SMS demand? Disrupted conventional PT service?). The transport analyst would thus need to apply other strategies to tackles such issues.

Computing accessibility of SMS raises new conceptual questions, which are new compared to more conventional modes. 
For conventional modes, there is a clear distinction between the \emph{realized} mobility and the \emph{potential for} mobility (with the aim of reaching opportunities). Accessibility captures the latter~\cite[p6]{Bertolini2019}, in that it represents what people \emph{can} reach, instead of what they \emph{have reached}. These two aspects are, however, entangled in SMS. Indeed, \rev{}{}{in} SMS\rev{}{}{, such as DRT or shared taxis,} vehicle trajectories are built to satisfy specific trip requests\rev{}{}{, and in systems such as bike or car sharing, vehicle availability is determined by the trajectories traveled by users moving such vehicles}. Such trajectories thus represent realized mobility, and computing accessibility based on them may cast doubt on whether the semantics of accessibility are still respected. To clarify this point, a peculiarity of SMS must be emphasized, compared to conventional transport modes. In conventional modes, the supply is given (e.g., road network, PT schedules) and the demand is realized by adapting to the supply. Accessibility is computed based on the supply, which thus ensures neatly assessing the potential for mobility. On the contrary, in SMS, supply is \rev{}{adapted}{impacted} in real time \rev{}{to}{by} the demand \rev{}{, and it}{ (in the sense described above for DRT and shared taxis or bike and car sharing). Supply} is thus \emph{demand-driven}. As a consequence, travel times via SMS are generally more favourable between locations that are visited by prevalent demand patterns and by specific socio-economic classes. This recently raised equity concerns, which are however out of our scope (\cite{Jiao2021,Brown2021,JinhuaZhaoPhDStudent2024Ch4}). In this context, in SMS the potential for mobility is tightly influenced by the realized mobility. As a consequence, accessibility is influenced by demand: SMS vehicles happen to prevalently be in locations of high demand, from which it is thus easier to visit more opportunities. Other locations may suffer from higher waiting times and detours, which translate to lower accessibility. However, this entanglement between potential for mobility and realized mobility is inherent in SMS and is not an artefact of our accessibility computation method. 

\rev{}{}{In this paper we applied our paper to DRT acting as a feeder to conventional PT. Our method can however be adapted to areas where the DRT can also provide door-to-door connectivity. In that case, instead of estimating travel times to/from hubs, they would be additionally estimated for different origin-destination pairs. Estimation would occur in a 4-dimensional space (latitude and longitude of the origin and of the destination), instead of the 2-dimensional space treated in this paper. Kriging, on which we based our estimation, is generalizable to $n$ dimensions, which ensures the applicability of our method also beyond feeder DRT services.}

The proposed method is showcased in a \emph{what-if} study \cite[Figure~9.1]{cascetta2009transportation}: ``\emph{What} is the impact on accessibility \emph{if} SMS is deployed?'' In other words, a given SMS deployment is given and its impact analysed. Alternatively, a \emph{what-to} method, to plan SMS in order to reduce the inequality in the accessibility distribution, has been presented by \cite{Wang2024}. However, they  only resort to theoretical models based on geometrical abstractions. Our method could allow replacing such abstract models with realistic estimations of accessibility, obtained via high-fidelity simulation models, so as to capture the peculiarities of the area under the study, e.g., road topology, demand and land-use patterns, etc.

Note that the goal of this paper is not to thoroughly study the accessibility in Paris-Saclay. The goal is rather to present a method to compute accessibility from SMS, and Paris-Saclay is just a case study where such a method is showcased. In consultancy work or case study-based work, where the interest is to understand the transport services of a particular region, one would consider the different sources of accessibility, i.e., to distinguish accessibility by car, bikes and, in general, by all the modes available. Our method would be used to provide one of such accessibilities, namely the one provided jointly by conventional PT and SMS. Such different accessibilities could be considered separately, for comparison reason, or aggregated in a single \emph{multimodal accessibility} indicator (\cite[(13)-(16)]{miller2020accessibility}). Such a detailed study lies outside the methodological scope of this paper. Along this direction, an interesting perspective would be to focus on low-demand areas, where accessibility is dominated by car, to understand whether car-based accessibility can be effectively replaced by SMS-based accessibility. Our hypothesis is that such a replacement can be successful only if (i)~conventional PT already exists in such areas and SMS can be deployed as a feeder and (ii)~density is not excessively low. \rev{aa}{We postulate that, in the other cases, cars remain irreplaceable for people to reach their opportunity. Our method can support future case-by-case investigation of such important questions.}{}

The concept of accessibility offers a succinct yet powerful representation of the combined effects of land use and transport services. Accessibility can be improved by acting on one or the other or both. In this paper, we only consider acting on transport, by introducing SMS. This modifies accessibility, through improving times~$T(\mathbf{u},\mathbf{u}',t)$ of~\eqref{eq:acc}. The impact of land-use is captured by the distribution of opportunities~$O_\mathbf u,\forall \mathbf u\in\pazocal C$. The method we proposed would allow a planner to evaluate, jointly, decisions related to the distribution of opportunities and the deployment of SMS, to achieve a desirable accessibility.

The proposed method could also be used within broader frameworks to evaluate the long-term impact of SMS on land-use and development. In Land Use and Transport Interaction (LUTI) models, accessibility indices (including isochrone-based ones, as in \cite{MeiChen2012,DeVries2022}) are employed as signals that can trigger relocation of households or businesses, determining urban evolution. A question may thus arise on whether improving accessibility in low-density areas via SMS might encourage urban sprawl. Our method could be employed to compute the accessibility indices fed into the LUTI models, in order to tackle the aforementioned question.

It must be noted that this work does not propose yet another accessibility computation tool. There are already several excellent and mature methods, tools and software that are available. Some of them became part of the toolkit of transport planners. Examples of tools are \cite{Bertolini2019,Conway2017,Conway2018, Byrd2023,biazzo2019accessibility,Miller2022}. This work proposes instead a statistical method that can be integrated with the pre-existing accessibility computation tools. Indeed, our method allows generating a schedule that represents SMS. Such a schedule is equivalent to SMS, in terms of travel times experienced by the users. Such a schedule is modelled as a time-expanded graph, which can then be given as input to the aforementioned tools, in the usual GTFS format. In this work, we picked one of such tools, namely CityChrone (\cite{biazzo2019accessibility}), for the simple reason that it is open source. But any other tool capable of reading GTFS could be used in its place.

%%%%%%%%%%%%%%%%%%%%%%%%%%%%%%%%%%%%%%%%%%%%%%%%%%%
%%%%%%%%%%%%%%%%%%% CONCLUSIONS %%%%%%%%%%%%%%%%%%%
%%%%%%%%%%%%%%%%%%%%%%%%%%%%%%%%%%%%%%%%%%%%%%%%%%%
\section{Conclusion}
\label{sec:conclusion}
We proposed the first method to compute the impact of SMS on accessibility, based on empirical observations of SMS trips. Our work is particularly relevant for transport authorities or operators that have deployed SMS or are planning to deploy them. Via our proposed method, such actors can quantitatively support claims such as ``if SMS is deployed in a certain zone, the number of opportunities that can be reached from that zone within 1h increases by~$50$ thousands'' (for instance, this exemplary value is visible in many hexagons of Figure~\ref{map:sc_diff_bc_drt_fd}). We believe that such quantitative arguments are extremely important to encourage the deployment of SMS in such communities, much more important than just estimating expected travel times or waiting times. In this sense, our work contributes to accelerating the adoption of SMS. On the contrary, in some cases, our method could be used to assess the uselessness of SMS in a certain region, if it does not considerably improve accessibility over the already deployed conventional PT, and can also prevent the emergence of an undesirable competition between conventional PT and SMS. Indeed, when deciding the areas where to deploy SMS or the size of the SMS fleet at a strategic level, planners and operators can run simulations and use our method to estimate the accessibility distribution corresponding to each configuration of the system. They could thus assess which configuration brings a desired improvement. Our method can also be used \emph{ex-post} to analyse how many additional opportunities people can reach thanks to an already deployed SMS, which can become a crucial argument to justify the sometimes high cost of the service.

We showcased our method on a dataset of SMS trips, obtained in a simulated future scenario in the Paris-Saclay area, inside the Paris Region. However, our method could be applied as is if real trips are available, instead of simulated ones. Since our method is data-driven, the quality of its output (accessibility results) strictly depends on the quality of its input (list of trips). Therefore, if transport operators are to use our method to obtain some findings related to accessibility, such findings can be trusted only if the information about SMS trips (no matter if simulated or real) can be trusted.

We pinpoint that the numerical results presented in this paper are not intended to give a detailed picture of how accessibility could change, specifically in the Paris-Saclay area, thanks to SMS. The results presented in this paper are just intended to show the types of findings achievable with our method. It is thus important to observe that, changing SMS dimensioning parameters, such as the fleet size, or other simulation parameters would change trip travel times. By applying our method, in the same way presented in this paper, one would thus obtain different accessibility estimations. Sensitivity of accessibility to simulation parameters is an interesting research perspective: by enabling accessibility assessment, our work opens the possibility to evaluate, in the future, how accessibility is affected by service dimensioning parameters and how they can be chosen so as to maximize accessibility improvement.

\begin{appendices}

\section{Appendices}
\subsection{Analysis of Temporal and Spatial Patterns of Shared Mobility Service trips}

\begin{figure}      
\center\includegraphics[width=100mm]{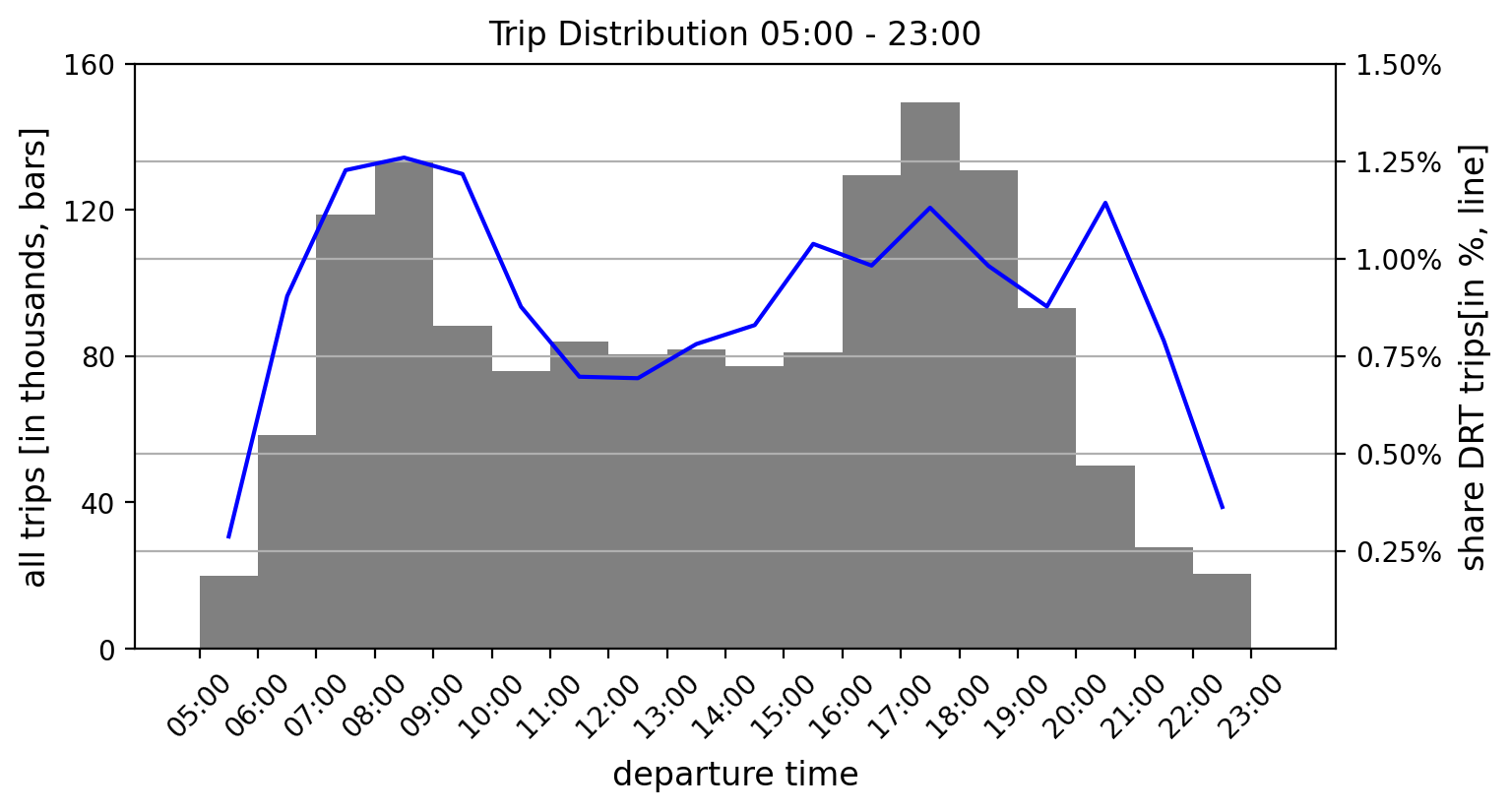}
\caption[Total Trips and SMS Trips over Time]{ Trips over time. One trip is defined by the departure within the study area of Paris-Saclay. A trip consisting out of multiple legs (e.g. walk + SMS + PT is considered as one trip)}
\label{fig:trips-tt-wt}
\end{figure}

In addition to the results about the accessibility estimation reported in the main text, it is useful to characterize the scenario from which such estimation is calculated. We thus report in this section some complementary information that gives a complete picture of the performance of the considered SMS.
Figure~\ref{fig:trips-tt-wt} clearly shows the morning peak $[7:00, 10:00[$, evening peak $[16:00, 19:00[$ and off-peak (all the other intervals).

%
%\[     [5:00, 7:00[, [7:00, 10:00[, [10:00, 16:00[, [16:00, 19:00[, [19:00, 23:00[ \]
%

\begin{figure}      
\center
\includegraphics[width=128mm]{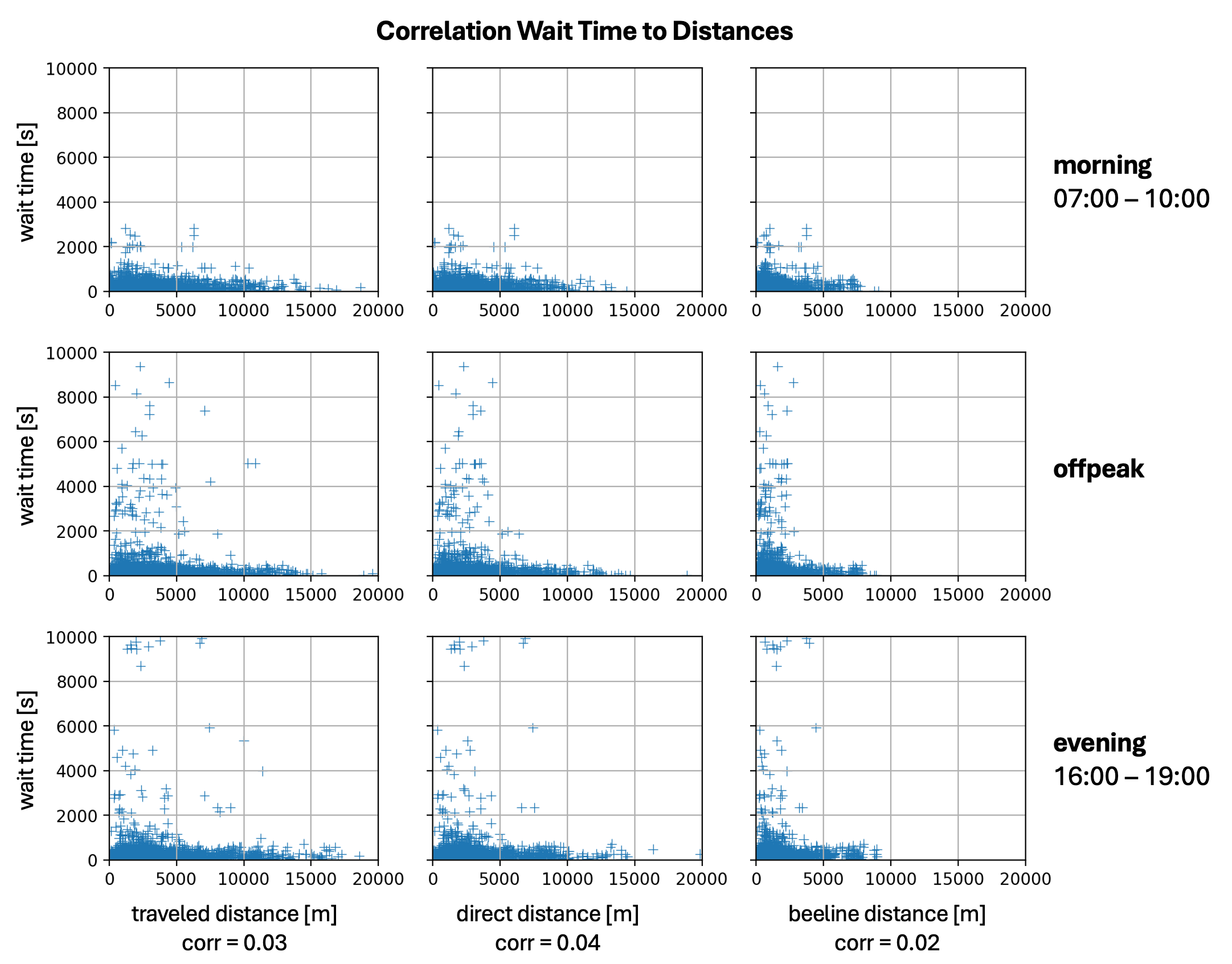}
\caption[Pair-plot Waiting Time - Travelled Distance, Direct Distance, Beeline Distance]{Relation between waiting time and distance measures.}
\label{fig:corr_pairplot_wt}
\end{figure}

\begin{figure}      
\center\includegraphics[width=120mm]{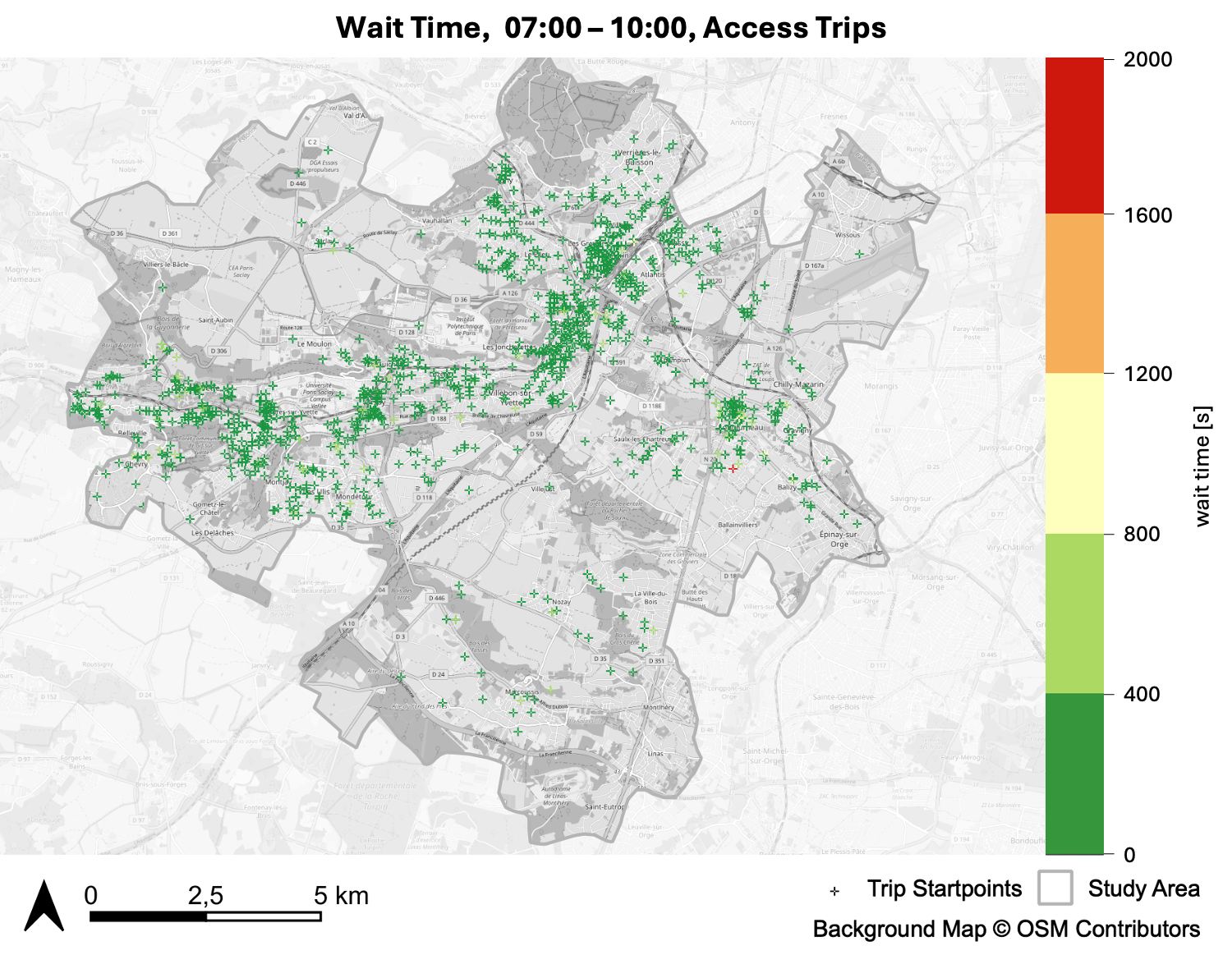}
\caption[Spatial Trend Waiting Time]{Spatial Trend of the Waiting Time,  in the interval [7:00 - 10:00]:  No clear pattern can be identified, indicating low spatial autocorrelation}
\label{map:wt_space}
\end{figure}

\begin{figure}      
\center\includegraphics[width=100mm]{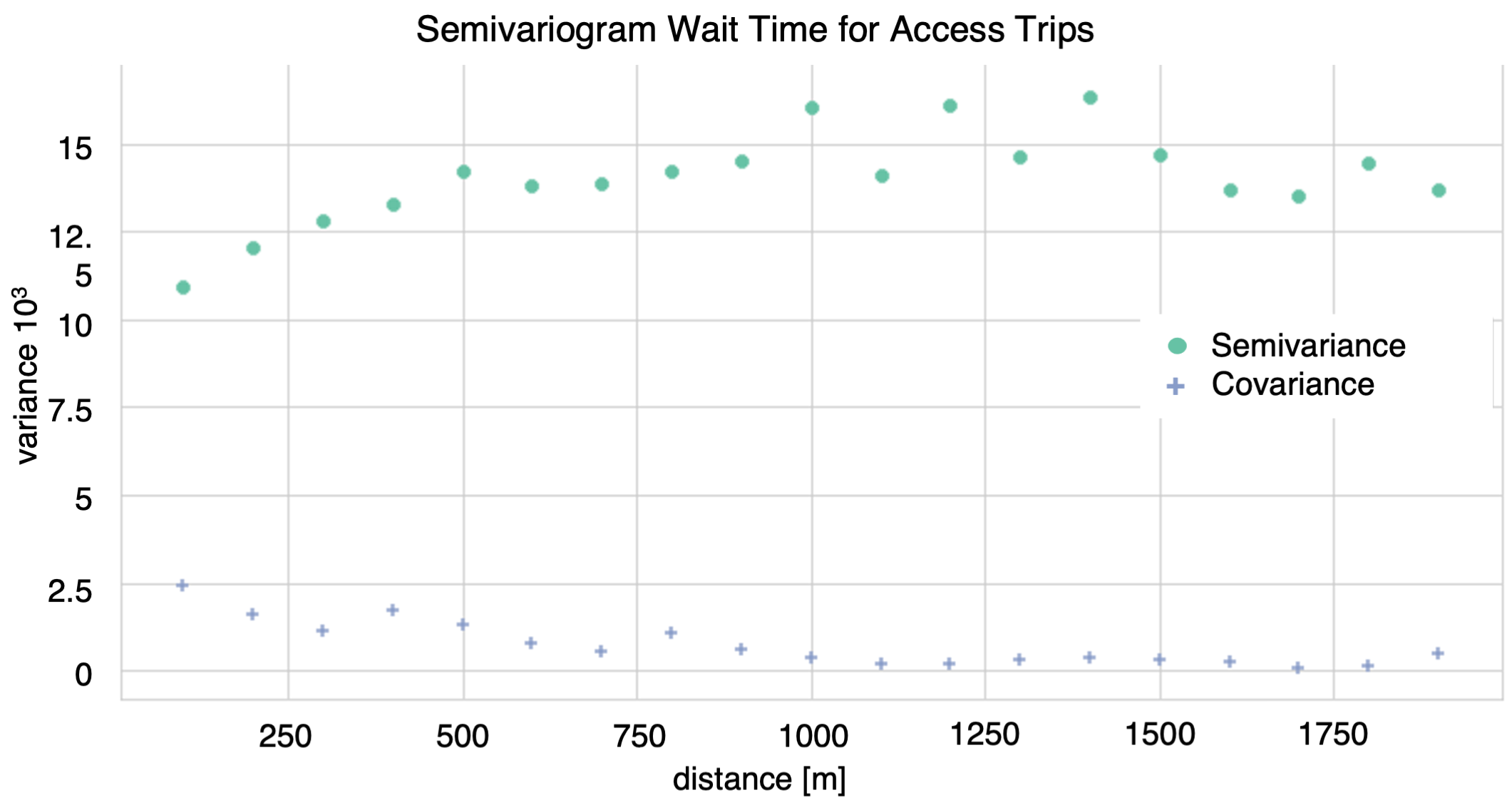}
\caption[Semivariogram Waiting Time of Access Trip Observations]{Spatial correlation of waiting time observations}
\label{semivar:wt}
\end{figure}

Figures~\ref{map:wt_space}-\ref{semivar:wt} depict the spatial autocorrelation of waiting times. Spatial autocorrelation of waiting times is less strong than for travel times, as commented in Sec.~\ref{sec:quality}. This is due to a weaker dependence of the waiting time on the distance (Figure~\ref{fig:corr_pairplot_wt}), compared with the case of travel times. However, similarity between observations still decays with their mutual distance (Figure~\ref{semivar:wt}), which is a favourable condition for applying Kriging. Excessive waiting times correspond to users that would in practice not be served. We record such times as a proxy for requests that are not easily handleable by the SMS service

\begin{figure}      
\center\includegraphics[width=100mm]{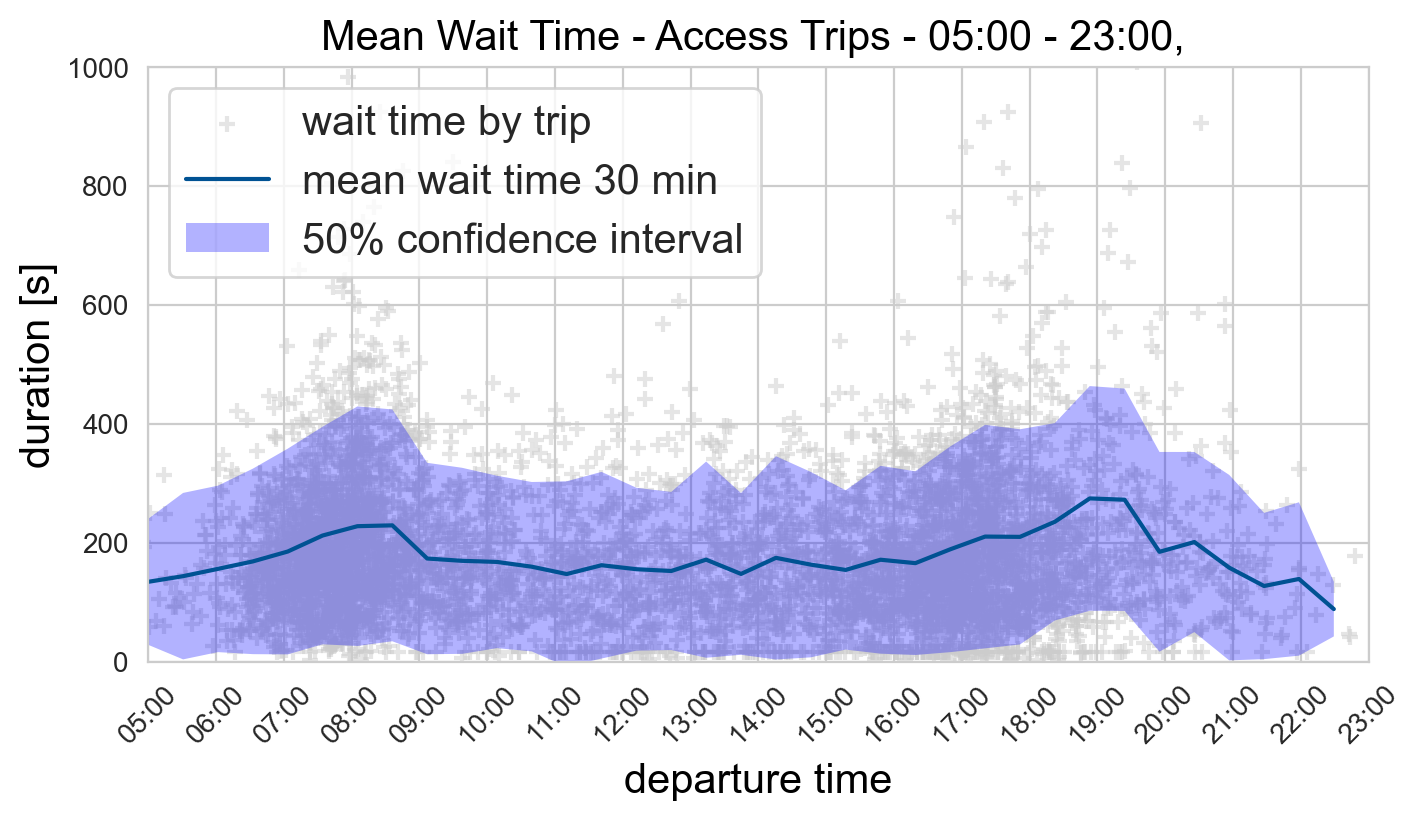}
\caption[Mean Waiting Time]{Mean Waiting Time: A moving average of waiting time during one day}
\label{fig:wt_mean}
\end{figure}

Figure~\ref{fig:wt_mean} shows that waiting time follows expected peak/off-peak patterns. Values are generally low, since the SMS fleet is sized so as to serve well the requests (see details in Section~\ref{Scenariodescribtion}).

\begin{figure}      
\center
\includegraphics[width=148mm]{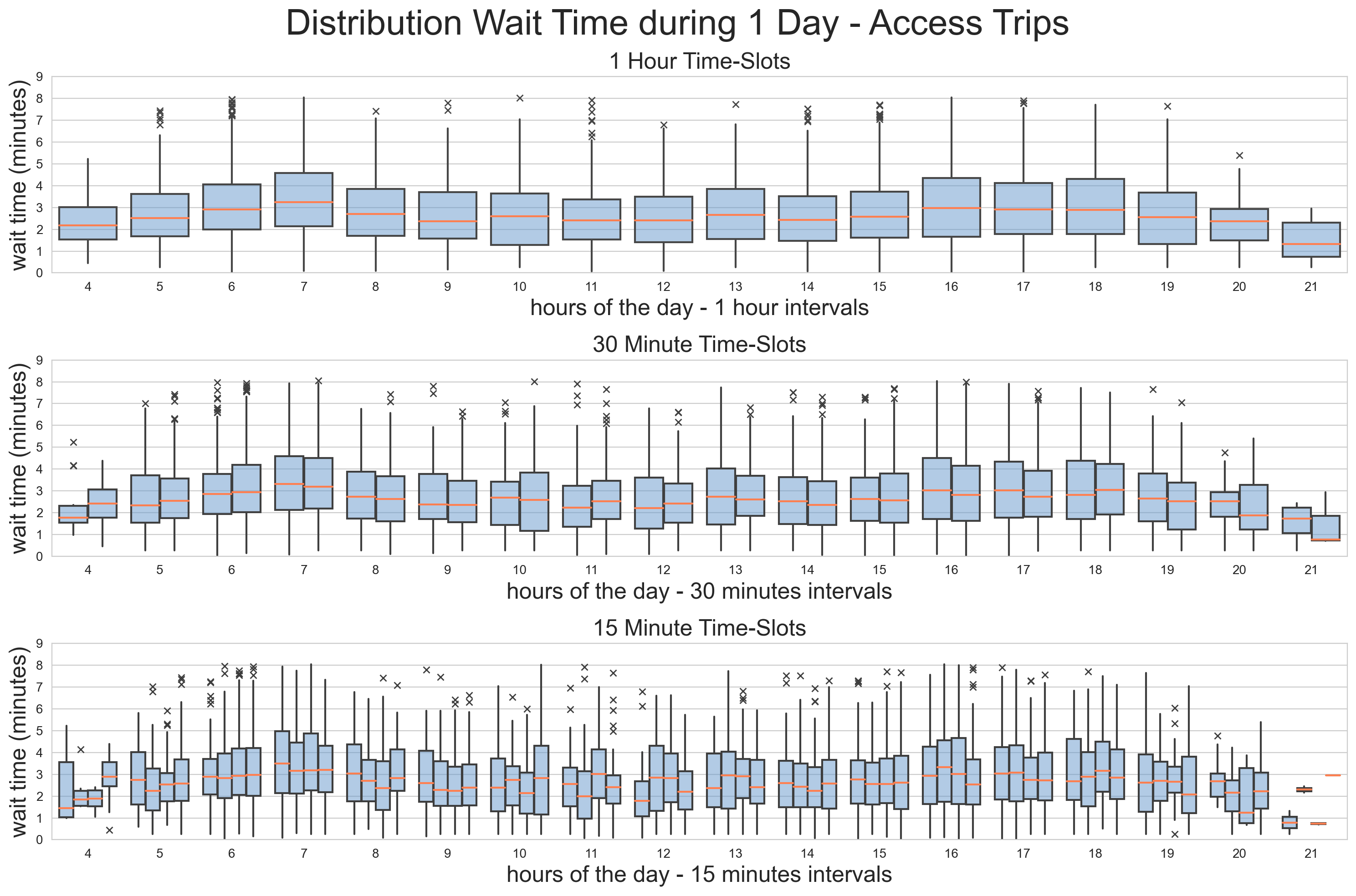}
\caption[Confidence and Preservation of Detail for Time Intervals]{Comparison of different timeslot sizes. Values exceeding 10 minutes are not depicted, as they are due to simulation events unpredictable for the SMS dispatcher}
\label{fig:distribution_wt_access}
\end{figure}
One hyperparameter we need to fix is the timeslot length (Section~\ref{sec:kriging}). Figure~\ref{fig:distribution_wt_access} shows that timeslots of 1h preserve the temporal pattern of trips, so 1h should be preferred to smaller timeslots, so as to perform Kriging within each timeslot with as many observations as possible.

\subsection{Procedure to derive weights~$\lambda_i$ of Ordinary Kriging}
\label{sec:derivation}
Let us assume we want to find estimation~$\hat w_{t_k}^\mathbf s(\mathbf x)$, i.e., the estimation of the waiting time for SMS to go from point~$\mathbf x$ to hub~$\mathbf s$ in timeslot~$t_k$. Based on~\eqref{eq:kriging-estimation}, this amounts to finding appropriate weights~$\lambda_i$.

The first step is to create an \emph{experimental semivariogram}, for which we have to choose hyperparameter $\Delta d$, called \emph{lag increment} (\cite[p72]{olea2000geostatistics}). This hyperparameter partitions the real axis into the following bins: $[n \Delta d- \Delta d/2, n \Delta d+ \Delta d/2[, n=0,1,2,\dots$. We follow the convention (\cite[p68-70]{oliver2007geostatistics}) of estimating the experimental semivariogram only at points~$n \Delta d$. Estimation $\hat \gamma(n \Delta d)$ is simply the average of values~$\gamma_{i,j}\equiv \frac{1}{2} \cdot (w_i - w_j)^2$ (see~\eqref{eq:gamma_ij}), only considering the points whose mutual distance is comprised between~$n \Delta d- \Delta d/2$ and~$n \Delta d+ \Delta d/2$. Hyperparameter~$\Delta d$ must not be too small, otherwise the experimental semivariogram could be erratic. It must not be too large either, otherwise it would fail to capture the variation of the variance for pairs of points varies over the distance. Such a hyperparameter is usually set with trial and error, coupled with visual inspection of the experimental semivariogram plot. After performing our analysis on our data, we found~$\Delta d=100m$ to be a good value, both for waiting time estimation and for in-vehicle travel time estimation (see Figures~\ref{semivar:tt}, \ref{semivar:wt}). 

The second step is to choose a model $\gamma(d):\mathbb R^+\rightarrow \mathbb R^+$, which is called \emph{theoretical semivariogram}, able to fit the set of values from the experimental semivariogram, where such a set is
$$\{(n\cdot \Delta d/2, \hat \gamma(n\cdot\Delta/2)) | n=0,1,2,\dots \}.$$
Various types of model types could be chosen and we adopt the usual Occam's razor criterion, namely ``choose the simplest model that describes sufficiently well the data'' (\cite[p77]{oliver2007geostatistics}). In our case, this is the \emph{bounded linear model} (\cite[(5.13)]{oliver2007geostatistics}). To fit the theoretical semivariance, one could use some automated procedure. But we follow the usual and more traditional approach of Kriging practitioners, by choosing the parameters of the model via some simple and intuitive criterion (\cite[Sec.~5.6]{oliver2007geostatistics}). In particular, only two parameters must be chosen in the bounded linear model. The first is the \emph{still variance}, which we estimated from the experimental semivariogram. The second is the \emph{range}, i.e. the distance starting from which the semivariogram reaches a \emph{plateau}, meaning that samples that are at a mutual distance larger than such a range, are basically independent. We empirically found that a range of~$3000m$ is appropriate for our dataset of observed trips.

Observe that the semivariograms (both experimental and theoretical) we are considering are called \emph{omnidirectional}, as they depend solely on the distance between pairs of points and not the direction.

The last step is to find the weights~$\lambda_i$ of~\eqref{eq:kriging-estimation}. In Ordinary Kriging, for any set of fixed weights~$\lambda_i$, estimation~$w_{t_k}^\mathbf s(\mathbf x)$ is intended as a random variable, whose  variance is equal to (see \cite[(8.2)]{oliver2007geostatistics}):

\[
\text{var}[w_{t_k}^\mathbf s(\mathbf x)]
=
2 \sum_{i\in\pazocal D_{t_k}^\mathbf s}\lambda_i\cdot \gamma(d(\mathbf x, \mathbf x_i))
-
\sum_{i,j\in\pazocal D_{t_k}^\mathbf s} \lambda_i\lambda_j \cdot  \gamma(d(\mathbf x_i, \mathbf x_j))
\]

The weights that will be used in~\eqref{eq:kriging-estimation} are those that minimize the variance above:
\begin{align}
\nonumber
\{\lambda_i^*\}_{i\in\pazocal D_{t_k}^\mathbf s}
&= \text{argmin}_{\{\lambda_i\}_{i\in\pazocal D_{t_k}^\mathbf s}}
\text{var}[w_{t_k}^\mathbf s(\mathbf x)]
\\
\nonumber
\text{s.t }\sum_{i\in\pazocal D_{t_k}^\mathbf s} \lambda_i &=1
\end{align}

\end{appendices}

\bibliographystyle{plain}
\bibliography{main.bib}

\begin{thebibliography}{10}

\bibitem{EUFacts2022}
{CO2 emissions from cars: facts and figures}.
\newblock Technical report, The European Parliament, 2022.

\bibitem{gtfsReference}
{GTFS Reference Document}, 2023.

\bibitem{Columbia}
Kriging interpolation, 2023.

\bibitem{abouelela2024we}
Mohamed Abouelela, David Dur{\'a}n-Rodas, and Constantinos Antoniou.
\newblock Do we all need shared e-scooters? an accessibility-centered spatial
  equity evaluation approach.
\newblock {\em Transportation Research Part A: Policy and Practice},
  181:103985, 2024.

\bibitem{MeiChen2012}
Meiwu An and Mei Chen.
\newblock An iterative approach to an integrated land use and transportation
  planning tool for small urban areas.
\newblock {\em Journal of Modern Transportation}, 20:160--167, 2012.

\bibitem{badeanlou2022ptanalysistool}
Amirhesam Badeanlou, Andrea Araldo, and Marco Diana.
\newblock Assessing transportation accessibility equity via open data.
\newblock {\em hEART}, 2022.

\bibitem{Brown2021}
Jesus~M. Barajas and Anne Brown.
\newblock {Not minding the gap: Does ride-hailing serve transit deserts?}
\newblock {\em Journal of Transport Geography}, (November 2020), 2021.

\bibitem{JaraDiaz2012waiting}
Leonardo~J Basso and Sergio~R Jara-D{\'\i}az.
\newblock Sec. 2.1 of integrating congestion pricing, transit subsidies and
  mode choice.
\newblock {\em Transportation Research Part A: Policy and Practice},
  46(6):890--900, 2012.

\bibitem{biazzo2019accessibility}
Indaco Biazzo, Bernardo Monechi, and Vittorio Loreto.
\newblock General scores for accessibility and inequality measures in urban
  areas.
\newblock {\em Royal Society open science}, 2019.

\bibitem{Calabro2021}
G.~Calabrò, A.~Araldo, M.~Ben-Akiva, et~al.
\newblock {Adaptive Transit Design: Optimizing Fixed and Demand Responsive
  Multi-Modal Transport via Continuous Approximation}.
\newblock In {\em {Transportation Research Part A}}, 2023.

\bibitem{carreyre2024demand}
F{\'e}lix Carreyre, Tarek Chouaki, Nicolas Coulombel, Ja{\^a}far Berrada,
  Laurent Bouillaut, and Sebastian H{\"o}rl.
\newblock On-demand autonomous vehicles in a periurban territory: a cost
  benefit analysis.
\newblock In {\em The 103rd Transportation Research Board (TRB) Annual
  Meeting}, 2024.

\bibitem{cascetta2009transportation}
Ennio Cascetta.
\newblock {\em Transportation systems analysis: models and applications}.
\newblock Springer Science \& Business Media, 2009.

\bibitem{Cats2022}
Oded Cats, Rafal Kucharski, Santosh~Rao Danda, and Menno Yap.
\newblock Beyond the dichotomy: How ride-hailing competes with and complements
  public transport.
\newblock {\em Plos one}, 17(1):e0262496, 2022.

\bibitem{chandra2013accessibility}
Shailesh Chandra, Muhammad~Ehsanul Bari, Prem~Chand Devarasetty, and Sharada
  Vadali.
\newblock Accessibility evaluations of feeder transit services.
\newblock {\em Transportation Research Part A}, 2013.

\bibitem{Chiles2018ok}
Jean-Paul Chil{\`e}s and Nicolas Desassis.
\newblock Fifty years of kriging.
\newblock In {\em {Handbook of Mathematical Geosciences}}, chapter~29. Springer
  International Publishing, 2018.

\bibitem{Chouaki2023}
Tarek Chouaki, Sebastian H{\"o}rl, and Jakob Puchinger.
\newblock Towards reproducible simulations of the grand paris express and
  on-demand feeder services.
\newblock In {\em TRB}, 2023.

\bibitem{Conway2017}
Matthew~Wigginton Conway, Andrew Byrd, and Marco van~der Linden.
\newblock Evidence-based transit and land use sketch planning using interactive
  accessibility methods on combined schedule and headway-based networks.
\newblock {\em Transportation Research Record}, 2653(1):45--53, 2017.

\bibitem{Conway2018}
Matthew~Wigginton Conway, Andrew Byrd, and Michael Van~Eggermond.
\newblock Accounting for uncertainty and variation in accessibility metrics for
  public transport sketch planning.
\newblock {\em Journal of Transport and Land Use}, 11(1):541--558, 2018.

\bibitem{craig2020gtfs}
Thomas Craig and Weston Shippy.
\newblock Gtfs flex--what is it and how is it used?
\newblock {\em National Center for Applied Transit Technology}, 2020.

\bibitem{Bossauw2019}
Nicola Da~Schio, Kobe Boussauw, and Joren Sansen.
\newblock Accessibility versus air pollution: A geography of externalities in
  the brussels agglomeration.
\newblock {\em Cities}, 84:178--189, 2019.

\bibitem{githubrepo}
Severin Diepolder.
\newblock Accessibility of dynamic transport, 2023.

\bibitem{Richter2021}
Norman Eppenberger and Maximilian~Alexander Richter.
\newblock The opportunity of shared autonomous vehicles to improve spatial
  equity in accessibility and socio-economic developments in european urban
  areas.
\newblock {\em European transport research review}, 13(1):32, 2021.

\bibitem{Erhardt2019a}
Gregory~D. Erhardt, Sneha Roy, Drew Cooper, Bhargava Sana, Mei Chen, and Joe
  Castiglione.
\newblock {Do transportation network companies decrease or increase
  congestion?}
\newblock {\em Science Advances}, (5), 2019.

\bibitem{EUTransport2020}
{EU Commission}.
\newblock {Mobility and transport}, 2020.

\bibitem{Fielbaum2024}
Andres Fielbaum and Javier Alonso-Mora.
\newblock Design of mixed fixed-flexible bus public transport networks by
  tracking the paths of on-demand vehicles.
\newblock {\em Transportation Research Part C: Emerging Technologies}, page
  104580, 2024.

\bibitem{fortin2016innovative}
Philippe Fortin, Catherine Morency, and Martin Tr{\'e}panier.
\newblock Innovative gtfs data application for transit network analysis using a
  graph-oriented method.
\newblock {\em Journal of Public Transportation}, 2016.

\bibitem{GeursVanWee2023}
Karst Geurs and Bert van Wee.
\newblock Accessibility: perspectives, measures and applications.
\newblock In {\em The transport system and transport policy: An introduction},
  pages 178--199. Edward Elgar, 2023.

\bibitem{GeursVanWee2013}
Karst~Teunis Geurs, Bert van Wee, and Piet Rietveld.
\newblock Discussing the logsum as an accessibility indicator.
\newblock In {\em XII NECTAR 2013 International Conference: Dynamics of Global
  and Local Networks}, 2013.

\bibitem{gow2024empirical}
Ian~D Gow and Tongqing Ding.
\newblock {\em Empirical research in accounting: Tools and methods}.
\newblock CRC Press, 2024.

\bibitem{Ettema2024}
Xiaodong Guan, Dea van Lierop, Zihao An, Eva Heinen, and Dick Ettema.
\newblock Shared micro-mobility and transport equity: A case study of three
  european countries.
\newblock {\em Cities}, 153:105298, 2024.

\bibitem{JinhuaZhaoPhDStudent2024Ch4}
Xiatong Guo.
\newblock {\em Towards A Robust Integrated Urban Mobility System: Public
  Transit and Ride-Sharing Systems - Chapter 4}.
\newblock PhD thesis, Massachusetts Institute of Technology (MIT), 2024.

\bibitem{Handcock1994}
Mark~S. Handcock and James~R. Wallis.
\newblock An approach to statistical spatial-temporal modeling of
  meteorological fields.
\newblock {\em Journal of the American Statistical Association}, (426), 1994.

\bibitem{Marshall2019}
Alejandro Henao and Wesley~E. Marshall.
\newblock {The impact of ride-hailing on vehicle miles traveled}.
\newblock {\em Transportation}, (6), 2019.

\bibitem{Miller2022}
Christopher~D Higgins, Yang~Luna Xi, Michael Widener, Matthew Palm, James
  Vaughan, Eric~J Miller, Amber DeJohn, and Steven Farber.
\newblock Calculating place-based transit accessibility.
\newblock {\em Journal of Transport and Land Use}, 15(1):95--116, 2022.

\bibitem{horl_synthetic_2021}
Sebastian Hörl and Milos Balac.
\newblock Synthetic population and travel demand for {Paris} and
  Île-de-{France} based on open and publicly available data.
\newblock {\em Transportation Research Part C: Emerging Technologies},
  130:103291, September 2021.

\bibitem{Srivastava1989}
Edward~H. Isaaks and R.~Mohan Srivastava.
\newblock {\em Applied Geostatistics}, chapter Chapter 12 ``Ordinary
  Kriging'''.
\newblock Oxford University Press, New York, 1989.

\bibitem{Jiao2021}
Junfeng Jiao and Fangru Wang.
\newblock {Shared mobility and transit-dependent population: A new equity
  opportunity or issue?}
\newblock {\em International Journal of Sustainable Transportation},
  15(4):294--305, 2021.

\bibitem{Laquidara2024}
Matt Laquidara.
\newblock A new kind of map for transit access.
\newblock \url{https://busgraphs.com/post/2024-10-03-segment-journeys/},
  October 2024.
\newblock Accessed: 2025-03-20.

\bibitem{hasif2022graph}
Cathia Le~Hasif, Andrea Araldo, Stefania Dumbrava, and Dimitri Watel.
\newblock A graph-database approach to assess the impact of demand-responsive
  services on public transit accessibility.
\newblock In {\em 15th ACM SIGSPATIAL International Workshop on Computational
  Transportation Science}, 2022.

\bibitem{Miller2020waiting}
Luyu Liu and Harvey~J Miller.
\newblock Does real-time transit information reduce waiting time? an empirical
  analysis.
\newblock {\em Transportation Research Part A: Policy and Practice},
  141:167--179, 2020.

\bibitem{VanWee2018}
Dimitris Milakis, Maarten Kroesen, and Bert {van Wee}.
\newblock Implications of automated vehicles for accessibility and location
  choices: Evidence from an expert-based experiment.
\newblock {\em Journal of Transport Geography}, 68:142--148, 2018.

\bibitem{miller2018accessibility}
Eric~J Miller.
\newblock Accessibility: measurement and application in transportation
  planning.
\newblock {\em Transport Reviews}, 2018.

\bibitem{miller2020accessibility}
Eric~J Miller.
\newblock Measuring accessibility: Methods and issues.
\newblock In {\em Roundtable on Accessibility and Transport Appraisal}.
  International Transport Forum, 2020.

\bibitem{molinski2022pyinterpolate}
Szymon Moli{\'n}ski.
\newblock Pyinterpolate: Spatial interpolation in python for point measurements
  and aggregated datasets.
\newblock {\em Journal of Open Source Software}, 2022.

\bibitem{nahmias2021drtaccessibility}
Bat-hen Nahmias-Biran, Jimi~B Oke, Nishant Kumar, Carlos Lima~Azevedo, and
  Moshe Ben-Akiva.
\newblock Evaluating the impacts of shared automated mobility on-demand
  services: An activity-based accessibility approach.
\newblock {\em Transportation}, 2021.

\bibitem{olea2000geostatistics}
Ricardo~A Olea.
\newblock Geostatistics for engineers and earth scientists, 2000.

\bibitem{Nielsen2018waiting}
Jens Parbo, Otto~A Nielsen, and Carlo~G Prato.
\newblock Sec. 4.2.2 of reducing passengers’ travel time by optimising
  stopping patterns in a large-scale network: A case-study in the copenhagen
  region.
\newblock {\em Transportation Research Part A: Policy and Practice},
  113:197--212, 2018.

\bibitem{Pottier2020}
Antonin Pottier, Emmanuel Combet, Jean-Michel Cayla, Simona de~Lauretis, and
  Franck Nadaud.
\newblock Qui emet du co2? panorama critique des inrgalites ecologiques en
  france.
\newblock {\em Revue de l'OFCE}, (5):73--132, 2020.

\bibitem{Boussauw2022}
Pedram Saeidizand, Koos Fransen, and Kobe Boussauw.
\newblock Revisiting car dependency: A worldwide analysis of car travel in
  global metropolitan areas.
\newblock {\em Cities}, 2022.

\bibitem{Geosciences}
B.~S.~Daya Sagar, Qiuming Cheng, and Frits Agterberg, editors.
\newblock {\em {Advances in Sensitivity Analysis of Uncertainty to Changes in
  Sampling Density When Modeling Spatially Correlated Attributes}}, chapter~19.
\newblock Springer Open, 2018.

\bibitem{GPE}
{SGP}.
\newblock { Fréquences prévisionnelles des trains aux heures de pointe du
  matin - Société du Grand Paris}, 2015.

\bibitem{Bertolini2019}
Cecilia Silva, Nuno Pinto, and Luca Bertolini.
\newblock {\em Designing accessibility instruments: Lessons on their usability
  for integrated land use and transport planning practices}.
\newblock Routledge, 2019.

\bibitem{Byrd2023}
Anson~F Stewart and Andrew~M Byrd.
\newblock Half-(head) way there: Comparing two methods to account for public
  transport waiting time in accessibility indicators.
\newblock {\em Environment and Planning B: Urban Analytics and City Science},
  50(8):2187--2202, 2023.

\bibitem{Vale2023}
David Vale and Andr{\'e}~Soares Lopes.
\newblock Accessibility inequality across europe: a comparison of 15-minute
  pedestrian accessibility in cities with 100,000 or more inhabitants.
\newblock {\em NPJ Urban Sustainability}, 3(1):55, 2023.

\bibitem{Wang2024}
Duo Wang, Andrea Araldo, and Mounim {El Yacoubi}.
\newblock Planning demand-responsive transit to reduce inequality of
  accessibility.
\newblock {\em Transportation Research Part A: Policy and Practice},
  199:104544, 2025.

\bibitem{DeVries2022}
Zhongqi Wang, Qi~Han, and Bauke De~Vries.
\newblock Land use spatial optimization using accessibility maps to integrate
  land use and transport in urban areas.
\newblock {\em Applied Spatial Analysis and Policy}, 15(4):1193--1217, 2022.

\bibitem{oliver2007geostatistics}
Richard Webster and Margaret~A Oliver.
\newblock {\em Geostatistics for environmental scientists}.
\newblock John Wiley \& Sons, 2007.

\bibitem{Bertolini2021}
J.~K. Wiersma, L.~Bertolini, and L.~Harms.
\newblock {Spatial conditions for car dependency in mid-sized European city
  regions}.
\newblock {\em European Planning Studies}, 29(7):1314--1330, 2021.

\bibitem{Szidarovsky1985}
S.~J. Yakowitz and F.~Szidarovsky.
\newblock A comparison of kriging with nonparametric regression methods.
\newblock {\em Journal of Multivariate Analysis}, 1985.

\bibitem{ZhangJie2018waiting}
Jie Zhang, David~ZW Wang, and Meng Meng.
\newblock Sec. 3.3.1 of which service is better on a linear travel corridor:
  Park \& ride or on-demand public bus?
\newblock {\em Transportation Research Part A: Policy and Practice},
  118:803--818, 2018.

\bibitem{zhou2021simulating}
Meng Zhou, Diem-Trinh Le, Duy~Quy Nguyen-Phuoc, P~Christopher Zegras, and
  Joseph Ferreira~Jr.
\newblock Simulating impacts of automated mobility-on-demand on accessibility
  and residential relocation.
\newblock {\em Cities}, 2021.

\bibitem{Zhu02052023}
Jiangtao Zhu, Ningke Xie, Zeen Cai, Wei Tang, and Xiqun (Michael)~Chen and.
\newblock A comprehensive review of shared mobility for sustainable
  transportation systems.
\newblock {\em International Journal of Sustainable Transportation},
  17(5):527--551, 2023.

\bibitem{Ziemke2023}
Dominik Ziemke and Joschka Bischoff.
\newblock Accessibilities by shared autonomous vehicles under different
  regulatory scenarios.
\newblock {\em Procedia Computer Science}, 220:747--754, 2023.

\bibitem{zwick_shifts_2022}
Felix Zwick, Nico Kuehnel, and Sebastian Hörl.
\newblock Shifts in perspective: {Operational} aspects in (non-)autonomous
  ride-pooling simulations.
\newblock {\em Transportation Research Part A: Policy and Practice},
  165:300--320, November 2022.

\end{thebibliography}

%%%%%%%%%%%%%%%%%%%%%%%%%%%%%%%%%%%%%%%%%%%%%%%%%%
%%%%%%%%%%%%%%%%%%%% ACK %%%%%%%%%%%%%%%%%%%%%%%%%
%%%%%%%%%%%%%%%%%%%%%%%%%%%%%%%%%%%%%%%%%%%%%%%%%%
\section*{Acknowledgements}

This work has been supported by The French ANR research project MuTAS (ANR-21-CE22-0025-01) and by BayFrance (FK-14-2021). It has also been supported by the European Union’s Horizon 2020 research and innovation programme under the Marie Skłodowska-Curie grant agreement no. 899987
%% IRT PART
Data were provided by the Anthropolis research project at the SystemX Technological Research
Institute, supported by the French government under the “France 2030” program. 

\section*{Author contributions statement}
A. A., S. D.,  C. A., S.M.: conceptualization\\
A.A.: theoretical development
\\
S. D.: code development, implementation, results
\\
S. D., A. A., S. M.: data analysis, writing, discussion
\\
T. C., S. H.: simulation \\
All authors reviewed the manuscript.

\end{document}